\documentclass[article,twocolumn,floats,nofootinbib,nobibnotes,longbibliography,superscriptaddress,10pt]{revtex4-1}

\usepackage{graphicx,amsmath,amssymb,hyperref}
\hypersetup{colorlinks=true, linkcolor=red, citecolor=magenta, urlcolor=blue}
\usepackage{dcolumn}
\usepackage{xcolor,colortbl}
\usepackage{enumitem}
\usepackage{eqnarray}
\usepackage{makecell}
\usepackage{cancel}
\usepackage{etoolbox}
\usepackage{comment}
\usepackage{amsfonts,mathtools}
\usepackage{tabularx,booktabs}
\usepackage{multirow}
\usepackage{rotating}

\definecolor{darkteal}{RGB}{0,121,150}

\begin{document}

\title{Joint 21-cm and CMB Forecasts for Constraining Self-Interacting Massive Neutrinos}

\newcommand\affSL{
\affiliation{Department of Physics, Ben-Gurion University of the Negev, Be'er Sheva 84105, Israel}
}

\newcommand\affEK{
\affiliation{Department of Physics, Ben-Gurion University of the Negev, Be'er Sheva 84105, Israel}
\affiliation{Texas Center for Cosmology and Astroparticle Physics, Weinberg Institute, Department of Physics, The University of Texas at Austin, Austin, TX 78712, USA}
}

\newcommand\affSG{
\affiliation{Texas Center for Cosmology and Astroparticle Physics, Weinberg Institute, Department of Physics, The University of Texas at Austin, Austin, TX 78712, USA}
}

\newcommand\affKB{
\affiliation{Texas Center for Cosmology and Astroparticle Physics, Weinberg Institute, Department of Physics, The University of Texas at Austin, Austin, TX 78712, USA}
}

\newcommand\affAR{
\affiliation{Dipartimento di Fisica e Astronomia ``Galileo Galilei'', Universit\` a di Padova, I-35131 Padova, Italy}
\affiliation{INFN Sezione di Padova, I-35131 Padova, Italy}
\affiliation{INAF-Osservatorio Astronomico di Padova, Italy}
}

\author{Sarah Libanore}
\email{libanore@bgu.ac.il}
\affSL

\author{Subhajit Ghosh}
\email{sghosh@utexas.edu}
\affSG

\author{Ely D. Kovetz}
\affEK

\author{Kimberly K. Boddy}
\affKB

\author{Alvise Raccanelli}
\affAR

\begin{abstract}
Self-interacting neutrinos provide an intriguing extension to the Standard Model, motivated by both particle physics and cosmology. Recent cosmological analyses suggest a bimodal posterior for the coupling strength $G_{\rm eff}$, favoring either strong or moderate interactions. These interactions modify the scale-dependence of the growth of cosmic structures, leaving distinct imprints on the matter power spectrum at small scales, $k\,>\,0.1\,{\rm Mpc}^{-1}$. For the first time, we explore how the 21-cm power spectrum from the cosmic dawn and the dark ages can constrain the properties of self-interacting, massive neutrinos. The effects of small-scale suppression and enhancement in the matter power spectrum caused by self-interacting neutrinos propagate to the halo mass function, shaping the abundance of small-- and intermediate--mass halos. It is precisely these halos that host the galaxies responsible for driving the evolution of the 21-cm signal during the cosmic dawn. We find that HERA at its design sensitivity can improve upon existing constraints on $G_{\rm eff}$ and be sensitive to small values of the coupling, beyond the reach of current and future CMB experiments. Crucially, we find that the combination of HERA and CMB-S4 can break parameter degeneracies, significantly improving the sensitivity to $G_{\rm eff}$ over either experiment alone. Finally, we investigate the prospects of probing neutrino properties with futuristic Lunar interferometers, accessing the astrophysics-free 21-cm power spectrum during the dark ages. The capability of probing small scales of these instruments will allow us to reach a percent-level constraint on the neutrino self-coupling.
\end{abstract}

\maketitle

\section{Introduction}\label{sec:intro}

\vspace*{-2pt}
Over the past decades, cosmology has made remarkable progress in mapping the large-scale structure (LSS) of the Universe and refining our understanding of fundamental physics. Observations of the cosmic microwave background (CMB) and large-scale galaxy surveys have provided stringent constraints on key cosmological parameters. These cosmological observations are sensitive to the fundamental interaction among particles, thus opening a window to search for physics beyond the Standard Model (SM). Further progress to scrutinize the concordance $\Lambda$CDM model and explore alternative cosmological scenarios requires new and complementary datasets.

Among the open questions in cosmology and fundamental physics, the properties of neutrinos remain elusive.
In the SM, there are three flavors of massless neutrinos. However, observations of neutrino oscillations indicate at least two of the three flavors have non-zero mass~\cite{Super-Kamiokande:1998kpq,Super-Kamiokande:2001bfk,Super-Kamiokande:2002ujc,Soudan2:2003qqa,Drexlin:2003fc,KamLAND:2004mhv,Super-Kamiokande:2005mbp,Messier:2006yg,DayaBay:2012fng}, but the absolute mass of the neutrino states and possible beyond-SM interactions remain unknown.

Using cosmological observations is a promising route to study the properties of neutrinos, due to the large abundance of neutrinos in the early Universe~\cite{Eisenstein:1997jh, Brandbyge:2010ge, Marulli:2011he, Castorina:2015bma,Gawiser:2000az,Doroshkevich:1980zs}.
Neutrinos affect the growth of structure differently at high and low redshifts due to their transition between the relativistic and non-relativistic regimes. The precise redshift when this transition occurs depends on the sum of neutrino masses $M_\nu$, which has been constrained, for example, using {\it Planck} 2018 CMB data~\cite{Planck:2018vyg} and baryon acoustic oscillations (BAO) results from the Dark Energy Survey Instrument (DESI)~\cite{DESI:2024mwx, Allali:2024aiv, DESI:2025zgx, DESI:2025ejh}. Currently, the {\it Planck} + DESI joint analysis places the strongest constraint on $M_\nu$~\cite{DESI:2025ejh}, beating even the design sensitivity of current and upcoming terrestrial experiments~\cite{KATRIN:2022ayy,Project8:2022wqh}. However, current observations are not yet sensitive enough to distinguish between different neutrino mass hierarchies~\cite{Herold:2024nvk}. 

Cosmological observations are also particularly sensitive to the free-streaming nature of neutrinos. In the SM, neutrinos free-stream after decoupling from the plasma around Big Bang Nucleosynthesis.
Preventing neutrinos from free-streaming enhances the CMB power spectra, modifies its acoustic phase-shift, and imprints characteristic features in the matter power spectrum~\cite{Cyr-Racine:2013jua,Archidiacono:2013dua,Forastieri:2017oma,Oldengott:2017fhy,Kreisch:2019yzn,Oldengott:2017fhy,RoyChoudhury:2020dmd,Ghosh:2019tab,Das:2020xke,Das:2023npl,Lancaster:2017ksf,Follin:2015hya,Baumann:2015rya,Montefalcone:2025unv}. A departure from the free-streaming nature would inevitably point towards the existence of beyond-SM physics in the neutrino sector.

An intriguing possibility is that the SM neutrinos experience self-interactions, which can be modeled by a 4-Fermi effective operator, induced by a mediator of mass $\mathcal{O}({\rm MeV})$~\cite{Cyr-Racine:2013jua}.
Recent studies have shown that measurements of the CMB and LSS show a moderate to strong preference for \emph{strong} neutrino self-interactions, almost six orders of magnitude stronger than SM weak interactions~\cite{Cyr-Racine:2013jua,Oldengott:2014qra,Oldengott:2017fhy,Kreisch:2019yzn,Das:2020xke,Das:2023npl,RoyChoudhury:2020dmd,Lancaster:2017ksf,Camarena:2023cku,Poudou:2025qcx,He:2023oke,He:2025jwp}
Moreover, these strong self-interaction scenarios have the potential to partially relax the Hubble and $S_8$ tensions~\cite{He:2025jwp,Poudou:2025qcx}.

Additionally, small-scale cosmological probes are crucial for studying self-interacting neutrinos across a wide range of interaction strengths.
Previous work studied the constraining power of the 21-cm power spectrum on interactions between neutrinos and dark matter~\cite{Dey:2022ini,Mosbech:2022uud} and of the 21-cm global signal on self-interactions between ultra-high-energy neutrinos~\cite{Dhuria:2024zwh}.
The effects at small scales may also be correlated with those at large scales, and a global analysis could strengthen constraints and break degeneracies with other cosmological and astrophysical parameters.
For example, high-redshift 21-cm data can be used in combination with CMB measurements to reduce the degeneracy between the optical depth $\tau$ and $M_\nu$ in CMB data, enabling the ability to constrain both parameters with good accuracy~\cite{Liu:2015txa,Shmueli:2023box,Facchinetti:2025hou}.

In this paper, we study how the high-redshift 21-cm power spectrum can probe the existence of self-interacting neutrinos.
Furthermore, we investigate the interplay between future CMB and 21-cm observations and their ability to place joint constraints on self-interaction coupling strength $G_{\rm eff}$ and the sum of neutrino masses $M_\nu$.
We perform forecasts for three different fiducial models with strong (SI$\nu$), moderate (MI$\nu$), and mild (mI$\nu$) neutrino self-interactions.
The SI$\nu$ model is based on the CMB results from Ref.~\cite{Camarena:2024daj}, which found a preference for self-interactions over $\Lambda$CDM.
The MI$\nu$ and mI$\nu$ models are ones beyond the reach of current CMB experiments.

Our main focus is the 21-cm power spectrum from the cosmic dawn, which probes small scales indirectly due to the relation between matter fluctuations and the abundance of the first galaxies.
We forecast for the design sensitivity of the Hydrogen Epoch of Reionization Array (HERA)~\cite{DeBoer:2016tnn}, which is a precursor of the Square Kilometer Array Observatory (SKAO)~\cite{Braun:2015zta,Braun:2019gdo} and already sets upper limits on the most extreme astrophysical models~\cite{HERA:2021bsv,HERA:2021noe,Lazare:2023jkg}.
We also forecast for futuristic observations of the 21-cm signal from the dark ages, which rely on advanced versions of SKAO or proposed Lunar detectors (e.g.,~Refs.~\cite{Silk:2020bsr,Chen:2024tvn,Ajith:2024mie,Burns:2021pkx,Chen:2020lok,Saliwanchik:2024nrp,Liu:2024tsg}).
Even if very challenging to access, the 21-cm signal from the dark ages would be extremely informative for cosmology, since it would be free from degeneracies with astrophysical sources, and it could be used to trace fluctuations on small scales that are still in the linear regime.
For the CMB, our forecasts use the specifications for the wide survey proposed for CMB-S4.

We find that the design sensitivity of HERA has much stronger constraining power on $G_{\rm eff}$ over CMB-S4 for the MI$\nu$ and mI$\nu$ models, since the CMB sensitivity is limited by Silk damping~\cite{Silk:1967kq}; for the SI$\nu$ model, CMB-S4 is the more sensitive probe.
In all cases, the combination of CMB-S4 and HERA outperforms either experiment alone, permitting $2\sigma$ constraints of $\sim 10\%$ on $\log_{10}G_{\rm eff}$.
For the SI$\nu$ model, including HERA improves upon the CMB-S4 constraint through its more precise determination of $\tau$.
For the mI$\nu$ model, CMB-S4 breaks the degeneracy in the 21-cm power spectrum between $M_\nu$ and small $G_{\rm eff}$, providing a remarkable improvement in the sensitivity to $G_{\rm eff}$ over the individual experiments.
We also extend our analysis to two configurations of Lunar detectors, and joint analyses with CMB-S4 achieves $2\sigma$ constraints of $\sim 1\%$--$5\%$ on $\log_{10}G_{\rm eff}$.

This paper is structured as follows. In Sec.~\ref{sec:nus} we summarize the properties of self-interacting neutrinos and review the effects of both standard and self-interacting neutrinos on the CMB and the matter power spectrum. In Sec.~\ref{sec:21cm} we introduce the 21-cm power spectrum and discuss its constraining power on the small-scale regime. 
In Sec.~\ref{sec:forecast} we introduce our forecasting formalism, describe our simulation setup, and summarize the properties assumed for the detectors included in our study. In Sec.~\ref{sec:results} we present our results for the CMB, 21-cm cosmic dawn, and 21-cm dark ages, as well as results from joint analyses between the CMB and the 21-cm signal.
We conclude in Sec.~\ref{sec:discussion}.

Throughout this paper, the effects of neutrino self-interaction on CMB and the matter power spectrum are calculated using \texttt{nuCLASS}\footnote{\url{https://github.com/subhajitghosh-phy/nuCLASS}}, which is a modified version of the Boltzmann solver \texttt{CLASS}~\cite{Blas:2011rf}\footnote{\url{https://github.com/lesgourg/class_public}}, and we have made our code publicly available.
The analysis of the 21-cm signal uses simulations produced with \texttt{21cmFirstCLASS}~\cite{Flitter:2023rzv,Flitter:2023mjj,Flitter:2024eay}\footnote{\url{https://github.com/jordanflitter/21cmFirstCLASS}}, which is an extension of the semi-analytical simulation code \texttt{21cmFAST} v3~\cite{Mesinger:2010ne}\footnote{\url{https://pypi.org/project/21cmFAST/}}.


\section{Cosmological effects of self-interacting neutrinos}\label{sec:nus}

Cosmological neutrinos play an important role in the growth of cosmic structures and have a significant impact on cosmological observables (e.g., see Refs.~\cite{Lesgourgues:2006nd,Dolgov:2002wy,Sakr:2022ans} for reviews).
At very high redshift, neutrinos are relativistic particles that are initially coupled to the photon-baryon fluid through weak interactions.
As the Universe expands, these interactions become inefficient, and neutrinos decouple during the radiation-dominated era at $T\sim 1\,{\rm MeV}$.
After decoupling, neutrinos free stream and have a relativistic energy density $\rho_\nu \simeq [1+7/8(4/11)^{4/3}N_{\rm eff}]\rho_\gamma$~\cite{Shvartsman:1969mm,Steigman:1977kc}, where $\rho_\gamma$ is the photon energy density and $N_{\rm eff}$ is the effective number of relativistic neutrino-like species.
Massive neutrinos eventually become non-relativistic due to the Hubble expansion and contribute instead as a matter component of the Universe.
However, they maintain a high thermal velocity that prevents them from clustering on small scales~\cite{Tremaine:1979we}. 

Incorporating self-interactions alters the standard cosmological impact of neutrinos.
The primary effect is the modification of their free-streaming nature: self-interactions drive neutrinos toward more fluid-like behavior.
There are well-motivated beyond-SM scenarios in which neutrinos possess self-interactions~\cite{Chacko:2003dt,Beacom:2004yd,Bell:2005dr,Berbig:2020wve,He:2020zns}, and we quantify the low-energy effect on cosmology with the effective operator
\begin{equation}
    \label{eq:Geff_operator}
    \mathcal{L} \supset {1 \over 2}G_{\rm eff}(\bar{\nu}\nu)(\bar{\nu}\nu) ,
\end{equation}
where $\nu$ is the left-handed neutrino and $G_{\rm eff}$ is a dimensionful coupling constant.
This interaction can be modeled as $\mathcal{L} \supset g_{\nu,ij} \varphi \bar{\nu}_i\nu_j$, where $\varphi$ is a scalar mediator and $i$ and $j$ are flavor indices.
For simplicity, we consider flavor-universal interactions and set the coupling $g_{\nu,ij} = g_\nu \delta_{ij}$ to be the same for all flavors with no cross-flavor interactions.
In this scenario, the effective coupling constant in ~Eq.~\eqref{eq:Geff_operator} is~\cite{Cyr-Racine:2013jua,Kreisch:2019yzn}
\begin{equation}
\label{eq:geffdef}
    G_{\rm eff} = \frac{|g_\nu|^2}{m_\varphi^2},
\end{equation}
where $m_\varphi$ is the mediator mass.
The neutrino self-interaction rate is $\Gamma \propto aG_{\rm eff}^2T_\nu^5$, where $a$ is the scale factor and $T_\nu$ is the neutrino temperature. This formulation has been extensively adopted in the literature to constrain the self-interaction coupling strength with cosmological observables~\cite{Cyr-Racine:2013jua,Oldengott:2014qra,Oldengott:2017fhy,Kreisch:2019yzn,Das:2020xke,Das:2023npl,RoyChoudhury:2020dmd,Lancaster:2017ksf,Camarena:2023cku,Poudou:2025qcx,He:2023oke,He:2025jwp}.

\begin{table}[t!]
    \centering
    \begin{tabular}{|c|c|c|c|c|}
    \hline
    & $\quad\Lambda$CDM$\quad$ & $\quad$SI$\nu$$\quad$ & $\quad$MI$\nu$$\quad$ & $\quad$mI$\nu$$\quad$ \\
    \hline
    $\log_{10}G_{\rm eff}$ & $-$ & $-$1.77 & $-$4 & $-$5 \\
    $h$ & 0.673 & 0.67 & 0.673 & 0.673 \\
    $\Omega_b$ & 0.0493 & 0.0498 & 0.0493  & 0.0493 \\
    $\Omega_c$ & 0.2612 & 0.2595 & 0.2612& 0.2612 \\
    $10^9 A_s$ & $2.093$ & $1.959 $ & $2.093$& $2.093$\\
    $n_s$ & 0.9637 & 0.9298 & 0.9637& 0.9637 \\
    $N_{\rm eff}$ & 3.01 & 2.82 & 3.01 & 3.01 \\
    $M_\nu$ & 0.051 & 0.08 & 0.051 & 0.051\\
    \hline
    {$\tau$} (Sec.~\ref{sec:nus} and \ref{sec:21cm}) & 0.052 & 0.052 & 0.052 & 0.052\\
    {$\tau$} (Sec.~\ref{sec:forecast}) & 0.052 & 0.045 & 0.054 & 0.053\\
    \hline
    \end{tabular}
    \caption{Fiducial values of the cosmological parameters for $\Lambda$CDM and the three models under consideration. These parameters are the neutrino self-interaction coupling strength $G_{\rm eff}$ (in ${\rm MeV}^{-2}$), Hubble parameter $h$, baryon energy density $\Omega_b$, cold dark matter energy density $\Omega_c$, primordial amplitude $A_s$, primordial spectral tilt $n_s$, effective number of relativistic neutrino-like species $N_{\rm eff}$, sum of neutrino masses $M_\nu$, and optical depth $\tau$. We use the top set of values for $\tau$ when describing the cosmological effects of self-interacting massive neutrinos in Secs.~\ref{sec:nus} and \ref{sec:21cm}. The bottom set of values are estimates from our {\tt 21cmFirstCLASS} simulations, assuming the fiducial II scenario astrophysical model in Table~\ref{tab:fiducial_astro}, and we use these values for our analysis in Sec.~\ref{sec:forecast}.}
    \label{tab:fiducial_cosmo}
\end{table}

Throughout this paper, we consider three fiducial models for massive neutrinos with strong (SI$\nu$), moderate (MI$\nu$), and mild (mI$\nu$) self-interactions.
The cosmological parameters we use for $\Lambda$CDM and for each self-interaction model are listed in Table~\ref{tab:fiducial_cosmo}.
The SI$\nu$ model uses the best-fit values from Table~V of Ref.~\cite{Camarena:2024daj} for the analysis of {\it Planck} 2018 TT+TE+EE data~\cite{Planck:2018vyg} combined with BAO measurements from 6dFGS~\cite{Beutler:2011hx}, SDSS MGS~\cite{Ross:2014qpa}, and BOSS DR12~\cite{BOSS:2016wmc}. However, note that we use the value of $N_{\rm eff}$ from Ref.~\cite{Camarena:2024daj} to define the degeneracy factor between the neutrino species (deg$_{\rm ncdm}$ in \texttt{nuCLASS}); converting back to $N_{\rm eff}$ produces the values we list in Table~\ref{tab:fiducial_cosmo}.

For the MI$\nu$ model, we use a value of $G_{\rm eff}$ that is close to the upper bound from Ref.~\cite{Camarena:2024daj}; for the mI$\nu$ model, we use a value of $G_{\rm eff}$ that is an order of magnitude smaller and is thus well beyond the reach of \textit{Planck}.
Since the MI$\nu$ and mI$\nu$ models are poorly constrained from CMB data alone, we fix the other parameters to their best-fit $\Lambda$CDM+$N_{\rm eff}$+$M_\nu$ values from Ref.~\cite{Camarena:2024daj}.
In all cases, we use the 95\% confidence level (CL) upper limit on $M_\nu$.

In the remainder of this section, we discuss how self-interacting massive neutrinos affect CMB anisotropies and the matter power spectrum, and we further motivate the choices for our fiducial models.
We provide details of the modified Boltzmann equations in Appendix~\ref{sec:modboltz}.


\subsection{Effects on the cosmic microwave background}\label{sec:nus_cmb}

Standard neutrinos impact the CMB through their gravitational effects.
Post neutrino decoupling, the relativistic, free-streaming neutrinos have a larger sound speed than the photon-baryon fluid, so the neutrinos gravitationally pull the photon-baryon fluid as their density perturbation modes oscillate.
This effect induces a phase shift of the CMB acoustic peaks towards larger scales (i.e., smaller multipoles $\ell$)~\cite{Baumann:2015rya,Bashinsky:2003tk}. Due to the free-streaming effects, neutrinos also diminish metric potentials (as we discuss in Sec.~\ref{sec:nus_pk}) and thus suppress the amplitude of the CMB power spectra $C_\ell$ at high multipoles. The magnitude of these effects depends on $N_{\rm eff}$ and $M_{\nu}$.

If neutrinos posses sufficient self-interactions after they decouple (from the photon-baryon fluid), they continue behaving as a fluid until their self-interactions decouple, thus delaying the onset of neutrino free streaming.
Only the scales that enter the horizon after self-decoupling experience the same gravitational effect from free-streaming neutrinos that is expected in $\Lambda$CDM.
Scales that enter the horizon before self-decoupling (and after neutrino decoupling) are not subject to the standard effects of neutrino free streaming; thus, on these scales, neutrino self-interactions induce an enhanced amplitude of CMB acoustic peaks and a shift towards larger multipoles, relative to $\Lambda$CDM.
The enhancement and shift are degenerate with $n_s$ and the angular size of the acoustic horizon $\theta_*$, respectively~\cite{Cyr-Racine:2013jua,Kreisch:2019yzn,Oldengott:2014qra,Das:2020xke}.

Because of the degeneracy between $G_{\rm eff}$ and $n_s$, previous CMB+LSS studies of self-interacting neutrinos have found a bimodal posterior for $\log_{10}G_{\rm eff}$~\cite{Cyr-Racine:2013jua,Oldengott:2014qra,Oldengott:2017fhy,Kreisch:2019yzn,Das:2020xke,Das:2023npl,RoyChoudhury:2020dmd,Lancaster:2017ksf,Camarena:2023cku,Poudou:2025qcx,He:2023oke,He:2025jwp}, with a tight posterior peak around $\log_{10}G_{\rm eff}\sim [-1.4,-1.8]$, thereby motivating our SI$\nu$ benchmark model.
We note that although we assume flavor-universal interactions, there is also a peak in $G_{\rm eff}$ around the same values for flavor-specific interactions~\cite{Das:2020xke,Das:2023npl}.
Since the value of $\theta_\ast$ (as well as $N_{\rm eff}$) shifts for these models, self-interacting neutrinos have been proposed as a possible solution to the Hubble tension~\cite{Kreisch:2019yzn,Das:2020xke}; however, subsequent studies~\cite{Bernal:2020vbb,RoyChoudhury:2020dmd,Camarena:2024daj,Poudou:2025qcx} have shown that including BAO measurements~\cite{Beutler:2011hx,Ross:2014qpa,BOSS:2016wmc} or CMB polarization data in the analysis significantly reduces the viability of resolving the tension.

\begin{figure}[t!]
    \centering
    \includegraphics[width=\columnwidth]{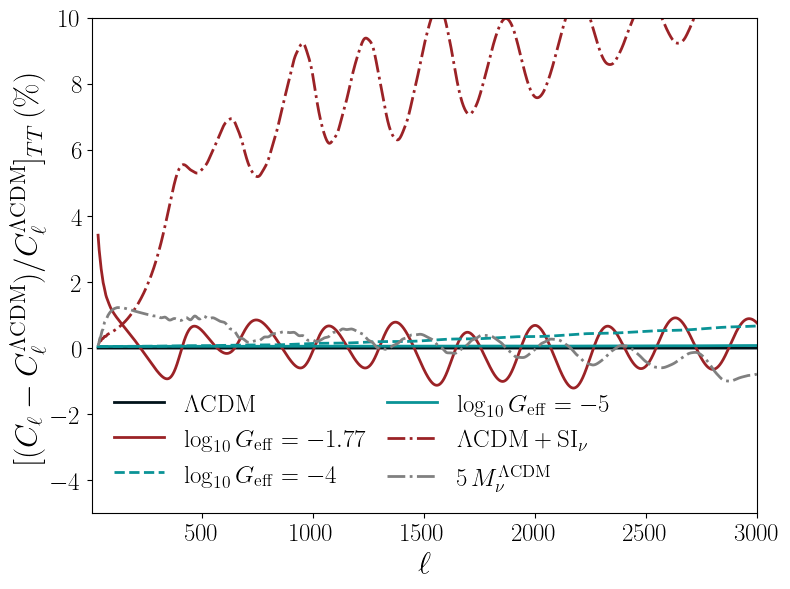}
    \caption{Residuals of the CMB temperature power spectrum for strong (SI$\nu$, solid red), moderate (MI$\nu$, dashed cyan), and mild (mI$\nu$, solid cyan) neutrino self-interactions with respect to $\Lambda$CDM. We use the fiducial values listed in Table~\ref{tab:fiducial_cosmo}. For comparison, the dot-dashed gray line illustrates the effect of increasing $M_\nu$ in $\Lambda$CDM{, while the red dot-dashed line represents the case in which neutrinos strongly interact, while other parameters are fixed to their $\Lambda$CDM values. This model is helpful to isolate the effect of SI$\nu$ on the observables, but is already ruled out by observations}.}
    \label{fig:Cl}
\end{figure}

Figure~\ref{fig:Cl} demonstrates the impact of self-interacting neutrinos on the CMB temperature power spectrum.
For SI$\nu$, free streaming is delayed up to a time around matter-radiation equality, leading to a net shift in the acoustic peak locations and their amplitudes. {If the cosmological parameters are fixed to their $\Lambda$CDM values, the resulting $C_\ell$ predictions from this model differ significantly from the \textit{Planck} data.  
In order to maintain a good fit, the changes due to neutrino strong self-interactions are partially compensated by changes to $A_s$, $n_s$, and $\tau$~\cite{Camarena:2024daj}.}

Moving away from the SI$\nu$ regime towards smaller values of $G_{\rm eff}$, previous studies have simply found upper limits, since the effects of self-interactions are pushed to smaller scales, corresponding to an earlier self-interaction decoupling time.
Our MI$\nu$ model corresponds to a value of $G_{\rm eff}$ near the upper limit in Ref.~\cite{Camarena:2024daj}, for which the scales of interest start to become more difficult to probe with the CMB due to Silk damping.
We choose a value of $G_{\rm eff}$ for the mI$\nu$ model that is an order of magnitude below that for the MI$\nu$, so detecting the impact of mI$\nu$ on the CMB would require a future instrument with sensitivity to smaller scales than \textit{Planck}.

In Fig.~\ref{fig:Cl}, the phase shift due to delayed free-streaming occurs on scales $\ell > 3000$, which is beyond the accessibility of \textit{Planck}, for the MI$\nu$ and mI$\nu$ models.
The only visible effect in the figure is the slightly larger amplitude at high multipoles, which does not produce a sufficient deviation from $\Lambda$CDM to be detected with \textit{Planck}.

As a point of comparison in Fig.~\ref{fig:Cl}, we show the effect of changing $M_\nu$ in a cosmology with no neutrino self-interactions, using the $\Lambda$CDM fiducial cosmology in Table~\ref{tab:fiducial_cosmo},
except we fix $\theta_\star$ instead of $h$ to its best-fit \textit{Planck} value and choose a larger sum of neutrino masses, set to five times that of the $\Lambda$CDM benchmark $(M_\nu = 5M_\nu^{\Lambda{\rm CDM}})$
An increase in the sum of neutrino masses causes a suppression in the temperature power spectrum at high multipoles, while an increase in the neutrino self-interaction strength (from mI$\nu$ to MI$\nu$) causes an enhancement.


\subsection{Effects on the linear matter power spectrum}\label{sec:nus_pk}

\begin{figure}[t!]
    \centering
    \includegraphics[width=\columnwidth]{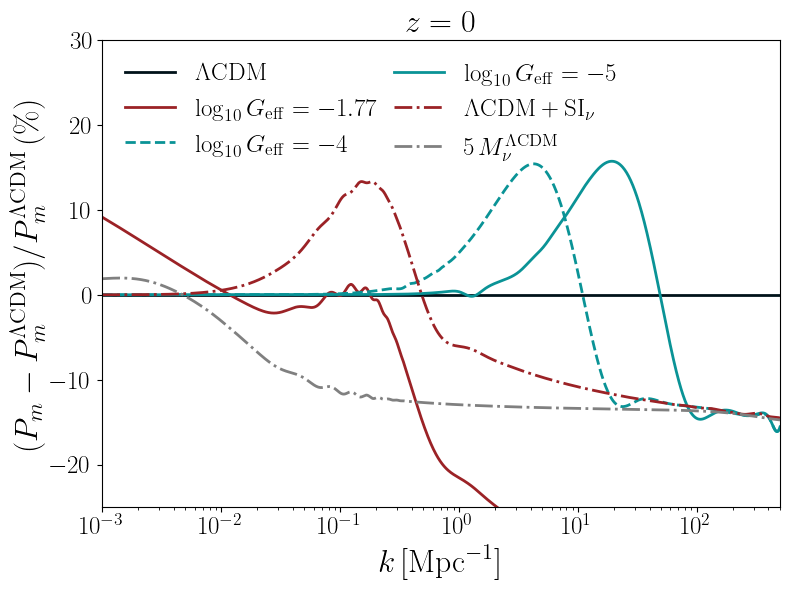}
    \caption{Residuals of the linear matter power spectrum at $z=0$. The legend is the same as in Fig.~\ref{fig:Cl}.}
    \label{fig:Pm}
\end{figure}

Massive neutrinos suppress matter clustering at scales that enter the horizon when neutrinos are relativistic. The suppression stems from two physical effects: relativistic neutrinos themselves do not cluster, and their radiation pressure hinders the clustering of cold dark matter and baryons, further suppressing structure formation.
Figure~\ref{fig:Pm} shows the additional suppression in the matter power spectrum for $M_\nu = 5M_\nu^{\Lambda{\rm CDM}}$.

The free-streaming nature of neutrinos also impacts the linear matter power spectrum $P_m(k,z)$.
Free-streaming induces a difference in the evolution of the potentials $\phi$ and $\psi$, which drive the growth of matter perturbations~\cite{Ma:1994dv,Bashinsky:2003tk}:
\begin{equation}\label{eq:pm_evol}
    d_c(k,t) = -\frac{9}{2}\phi_P+k^2\int dt'\,t' \psi(k,t')\ln\left(\frac{t'}{t}\right),
\end{equation}
where $d_c = \delta_c -3\phi$, $\delta_c$ is the dark matter density contrast in Newtonian gauge ($d_c=\delta_c$ at late times), $\phi_P$ is the primordial value of $\phi$ on large scales, and $t$ is conformal time. In particular, $\psi(k,t)$ decays for modes that enter the horizon during the radiation-dominated era.
In the presence of free-streaming neutrinos, the anisotropic stress is $\phi-\psi\neq 0$, which results in a smaller superhorizon initial value of $\psi$ and thus a faster decay~\cite{Lesgourgues:2006nd}.

Incorporating neutrino self-interactions suppresses anisotropic stress before the onset of neutrino free streaming. As a result, there is a scale-dependent impact on $P_m(k,z)$, as shown in Fig.~\ref{fig:Pm} for $z=0$. Small-scale modes that enter the horizon while neutrinos are still tightly self-coupled experience an anisotropic stress $\phi-\psi\sim 0$; therefore, the initial value of $\psi$ is higher, but its decay is slower. The overall effect is the suppressed growth of $d_c$, relative to $\Lambda$CDM. Neutrinos also affect the scale dependence of the galaxy bias (see e.g.,~\cite{Raccanelli:2017kht}), which we do not include in this work. 
In contrast, modes that enter the horizon close to neutrino self-decoupling experience a boost in the initial value of $\psi$, while subsequent evolution returns it to its $\Lambda$CDM behavior. Thus, the amplitude of density perturbations for these modes is strongly boosted. Finally, large-scale modes that enter the horizon after neutrino self-decoupling remain unchanged with respect to $\Lambda$CDM.

\begin{figure*}[ht!]
    \centering
    \includegraphics[width=2\columnwidth]{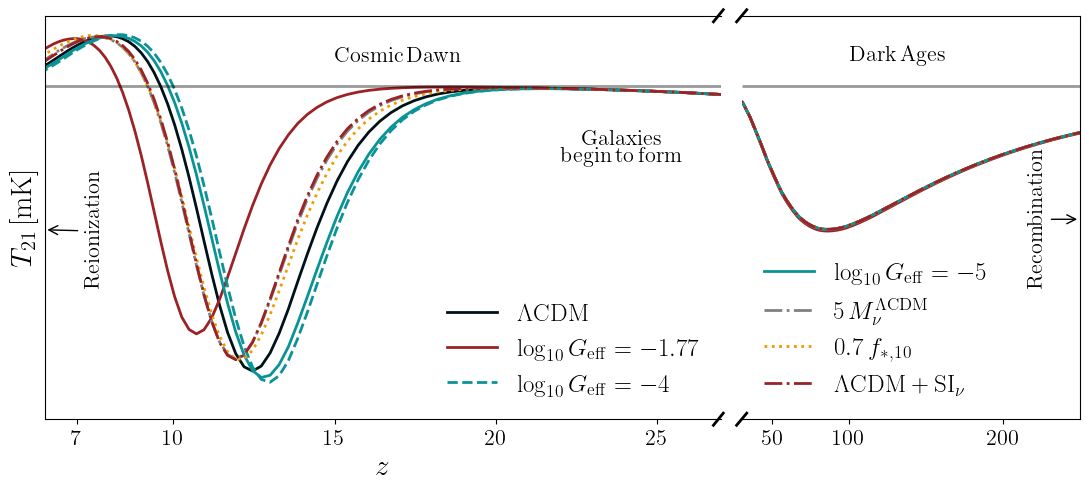}
    \caption{Illustrative plot of the 21-cm global signal across cosmic history. The evolution of $T_{21}(z)$ is estimated from the \texttt{21cmFirstCLASS} simulation, which evolves the 21-cm signal consistently with the chosen cosmological and astrophysical models, from recombination ($z\sim 1100$) down to reionization ($z\sim 6$). The legend is the same as in Fig.~\ref{fig:Cl}, but we also show the effect of reducing the star formation efficiency $f_{*,10}$ from its baseline. The horizontal line shows $T_{21}(z)=T_{\rm CMB}$.}
    \label{fig:global_signal}
\end{figure*}

Figure~\ref{fig:Pm} highlights the impact of neutrino interactions on the matter power spectrum, with respect to $\Lambda$CDM. The scale that enters the horizon around neutrino self-decoupling experiences the largest boost in $P_m(k)$. For the MI$\nu$ and mI$\nu$ models, the boost occurs on scales $k\sim 1-100\,{\rm Mpc}^{-1}$, and there is a moderate $\sim 10\%$ suppression at larger $k$. For the SI$\nu$ model, the shape of $P_m(k)$ with respect to $\Lambda$CDM arises from two effects. First, there is a boost $k\sim 0.1\,{\rm Mpc}^{-1}$ caused by the neutrino self-interactions{, which can be observed both when the other parameters are set to their $\Lambda$CDM values and when they are set to Tab.~\ref{tab:fiducial_cosmo}. In this second case}, the primordial spectral tilt for the SI$\nu$ model is smaller than the tilt for $\Lambda$CDM, as listed in Table~\ref{tab:fiducial_cosmo}. The smaller tilt results in the relative enhancement of $P_m(k)$ on very large scales, as well as the steep suppression for $k\gtrsim 1\,{\rm Mpc}^{-1}$.

The scale-dependent nature of the deviation from $\Lambda$CDM in all models is crucial for probing them using small-scale observations.
The large-$k$ suppression is similar in magnitude to the suppression from additional neutrino mass, controlled by the value $M_\nu$.
However, the notable enhancement of the matter power spectrum for the moderate and mild self-interaction models provides a distinct feature that can help distinguish these models.


\section{Impact of self-interacting neutrinos on the 21-cm signal}\label{sec:21cm}

At redshift $z\gtrsim 6$, the 21-cm line is sourced by neutral hydrogen (H) in the intergalactic medium (IGM). To study its signal, the main observable quantities are the brightness temperature 
\begin{equation}\label{eq:T21}
    T_{21}(z) = \frac{T_s(z)-T_{\rm CMB}(z)}{1+z}(1-e^{-\tau_{21}(z)}),
\end{equation}
and its spatial fluctuations $\delta T_{21}$, which are used to construct the 21-cm power spectrum
\begin{equation}\label{eq:Delta_21}
    \Delta_{21}^2({k},z) = \frac{k^3}{2\pi^2}\delta^D(\boldsymbol{k}-\boldsymbol{k}')\langle\delta T_{21}(\boldsymbol{k},z)\delta T_{21}^*(\boldsymbol{k}',z)\rangle.
\end{equation}

The evolution of $T_{21}(z)$ in Eq.~\eqref{eq:T21} follows the evolution of the difference between the CMB temperature $T_{\rm CMB}$ and the spin temperature $T_s(z)$, which provides an estimate of the relative number densities of H in the triplet and singlet states. In turn, the evolution of $T_s(z)$ is driven by the coexistence of different processes that excite H in the IGM, coupling its value either with the CMB temperature or with the gas temperature $T_{\rm gas}(z)$. 
The 21-cm optical depth is~\cite{Barkana:2000fd}
\begin{equation}
    \tau_{21}=(1+\delta_m)x_{\rm HI}\frac{T_0}{T_s}\frac{H(z)}{\partial v_r}(1+z),
\end{equation}
where $T_0$ is a normalization factor, $x_{\rm HI}$ is the neutral hydrogen fraction, $\delta_m$ is the matter density contrast, $H(z)$ the Hubble parameter, and $\partial v_r$ the line-of-sight gradient of the velocity. We summarize the key expressions required to model the evolution of all these quantities in Appendix~\ref{sec:app_21cm}.

Based on the dominant coupling process, we can distinguish between two different eras: 
\begin{itemize}
    \item the dark ages ($z\sim [30,200]$), during which $T_{\rm gas}$ cools adiabatically and  $T_s$ is coupled to its value via collisional coupling between the H atoms; and
    \item cosmic dawn ($z\sim [6,30]$), during which radiation sourced by stars and galaxies couples $T_s$ to $T_{\rm gas}$, while at the same time heating the gas and increasing $T_{\rm gas}$. The cosmic dawn is followed by the epoch of reionization ($z\sim [4,6]$), in which UV radiation from astrophysical sources ionizes the gas, thus decreasing the H optical depth $\tau_{21}$ and damping the 21-cm signal until it disappears.
\end{itemize}

The amplitude of the 21-cm signal is strongly affected by the choice of the underlying cosmological and astrophysical models. Extensive reviews can be found in Refs.~\cite{Barkana:2000fd,Furlanetto:2006jb,Pritchard:2011xb} and references therein; we simply visualize the evolution of the average value of $T_{21}(z)$, namely the global signal, in Fig.~\ref{fig:global_signal}. We obtain these various $T_{21}(z)$ curves by running \texttt{21cmFirstCLASS} under different choices of cosmological and astrophysical parameters, evolving the 21-cm signal consistently from recombination to reionization. 

Figure~\ref{fig:global_signal} highlights how different neutrino properties can speed up or slow down the evolution of the global 21-cm signal during cosmic dawn. However, it also shows how such changes can be degenerate with the choice of astrophysical parameters, such as the star formation efficiency $f_{*,10}$, which we discuss in Sec.~\ref{sec:astro}. Since different neutrino properties introduce scale-dependent effects in the matter power spectrum, we expect to be able to break degeneracies with the astrophysical parameters by looking at the 21-cm power spectrum $\Delta_{21}^2({k},z)$ instead. Using $\Delta_{21}^2({k},z)$ is also crucial to detect signatures of neutrino properties during the dark ages; here, in fact, the global signal is blind to such models, since they do not introduce extra energy injection in the IGM compared to $\Lambda$CDM.

In the following, we recap the key aspects required to understand how the amplitude of the 21-cm signal evolves and how $\Delta_{21}^2(k,z)$ traces the shape of the underlying $P_m(k,z)$, highlighting why we expect it to be sensitive to the small scales affected by the self-interacting neutrino models. 


\begin{figure*}[t!]
    \centering
    \includegraphics[width=2\columnwidth]{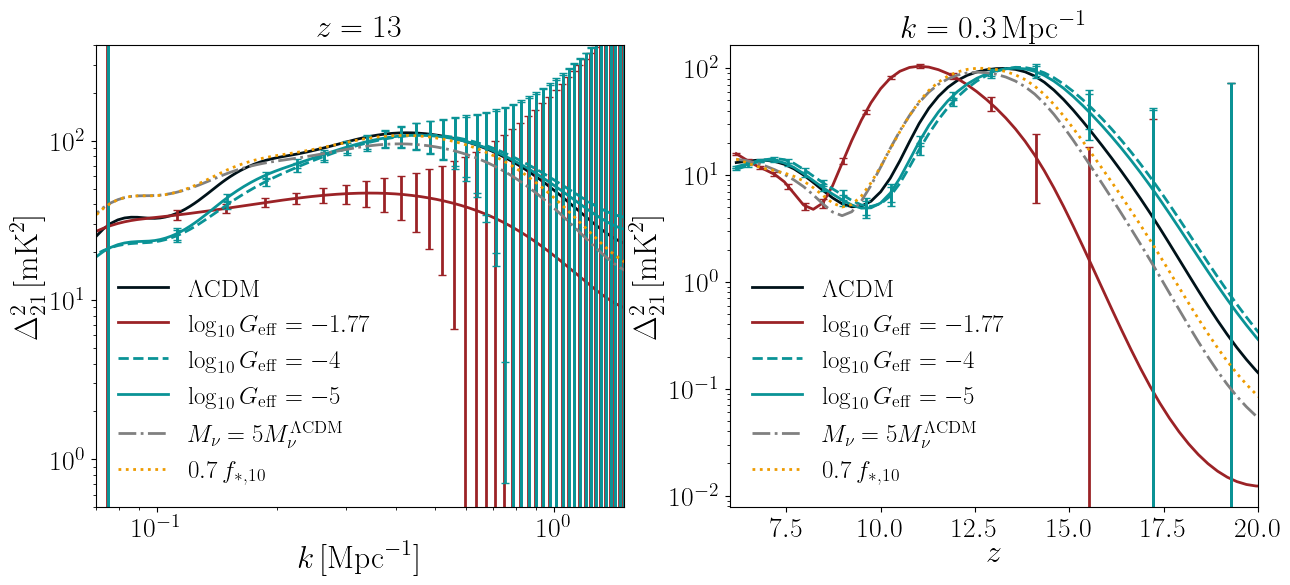}
    \caption{21-cm power spectrum during cosmic dawn for fixed $z$ (left) and for fixed $k$ (right). The legend is the same as in Fig.~\ref{fig:Cl}, {but here we did not include the  $\Lambda$CDM+SI$_\nu$ model}. We show $z = 13$, which coincides with the redshift of the 21-cm absorption peaks in Fig.~\ref{fig:global_signal}. We also show $k=0.3\,{\rm Mpc}^{-1}$, since it is in the range of scales already probed by HERA Phase I~\cite{HERA:2021bsv,HERA:2021noe,Lazare:2023jkg}. The errorbars represent the design sensitivity of the HERA detector we adopt for our analysis in Sec.~\ref{sec:forecast_detector}.}
    \label{fig:T21_pk_CD}
\end{figure*}

\vspace*{-.2cm}
\subsection{Cosmic dawn}\label{sec:21cm_CD}

When the first galaxies in the Universe begin to form, their radiation has enormous consequences for the conditions of the IGM, as it alters the status of the atoms in the gas, increases the gas temperature, and ionizes neutral H. 
The coupling between $T_s$ and $T_{\rm gas}$ during this epoch is caused by the Lyman-$\alpha$ (Ly$\alpha$) flux emitted by the stars, which is easily absorbed by H through the Wouthuysen-Field effect~\cite{Wouthuysen:1952AJ.....57R..31W,Field:1958PIRE...46..240F}. At the same time, the X-ray component of the radiation heats the gas~\cite{Oh:2000zx,Ricotti:2003vd,Pritchard:2006sq}, and the ability of the UV component to escape into the IGM turns on the epoch of reionization~\cite{Mesinger:2010ne,Ghara:2024xri}.

As we describe in Secs.~\ref{sec:astro} and~\ref{sec:hmf}, the detailed evolution of $T_{21}(z)$ and its fluctuations during cosmic dawn depends both on the shape of the underlying matter power spectrum and on the model of star formation. Therefore, the choice of cosmology and astrophysics impacts the shape of the expected 21-cm power spectrum during the cosmic dawn. 
As an example, in Fig.~\ref{fig:T21_pk_CD}, we show the $\Delta_{21}^2(k,z)$ produced by our \texttt{21cmFirstCLASS} simulations.
We compare the neutrino self-interaction models with $\Lambda$CDM at fixed $z$ and at fixed $k$, highlighting the features that each neutrino model introduces, which may be distinguishable with the sensitivity of upcoming 21-cm detectors. Moreover, the figure demonstrates that by using $\Delta_{21}^2(k,z)$, we can break degeneracies between astrophysical and cosmological parameters, due to the scale-dependent effect from cosmology.

\vspace*{-.2cm}
\subsubsection{Halo mass function}\label{sec:hmf}
\vspace*{-.2cm}

The key ingredient to model the evolution of $\Delta^2_{21}(k,z)$ is the star formation rate density (SFRD) $\dot{\rho}_*(\boldsymbol{x},z)$. This determines the intensity of the Ly$\alpha$ radiation that couples $T_s$ to $T_k$, as well as the intensity of the X-ray heating. A summary of the key quantities required to model these processes can be found in Appendix~\ref{sec:app_21cm}.
Following Refs.~\cite{Park:2018ljd,Munoz:2021psm}, we model the SFRD as a function of the star formation rate (SFR) $\dot{M}_*$, which depends on the host dark matter halo mass $M_h$, convolved with the local halo mass function (HMF) $dn/dM_h(\boldsymbol{x},z)$:
\begin{equation}\label{eq:sfrd}
    \dot{\rho}_*(\boldsymbol{x},z) =  \int_{M_{\rm min}}^\infty dM_h\frac{dn}{dM_h}(\boldsymbol{x},z)\dot{M}_*(M_h,z)f_{\rm duty}(M_h),
\end{equation}
where $f_{\rm duty}(M_h)$ is the duty cycle, which suppresses the SFR below a certain turnover mass, $M_{\rm turn}$. We set $M_{\rm min}=10^5\,M_\odot$, but this choice does not affect the final result, as long as $M_{\rm min}<M_{\rm turn}$.

{\color{black} We can rewrite Eq.~\eqref{eq:sfrd} in terms of the local collapse fraction of matter into halos with mass $M_h \geq M_{\rm min}$
\begin{equation}\label{eq:fcoll}
    f_{\rm coll}(\boldsymbol{x},z,M_{\rm min}) = \frac{1}{\bar{\rho}_{m,0}}\int_{M_{\rm min}}^\infty dM_h\frac{dn}{dM_h}(\boldsymbol{x},z)M_h, 
\end{equation}
where $\bar{\rho}_{m,0}$ is the mean matter density at $z=0$. As we discuss in detail in Sec.~\ref{sec:astro}, the star formation rate can be approximated as a power law of the host halo mass, $\dot{M_*}(M_h)=A_{\rm SFR}M_h^{\alpha_*}/\bar{\rho}_m$, where the $A_{\rm SFR}$ factor is independent of $M_h$, and $\alpha_*$ depends on the model. 
Since
\begin{equation}
    \frac{df_{\rm coll}}{dM_h}(\boldsymbol{x},z) = -\frac{M_h}{\bar{\rho}_m}\frac{dn}{dM_h},
\end{equation}
we can define
\begin{equation}\label{eq:hmf_to_fcoll}
    \mathcal{F}_{\rm coll}(\boldsymbol{x},z)= -\int d\log M_h \frac{df_{\rm coll}}{dM_h}(\boldsymbol{x},z)M_h^{\alpha_*}f_{\rm duty}(M_h),
\end{equation}
so that the SFRD can be re-expressed as 
\begin{equation}
    \dot{\rho}_*(\boldsymbol{x},z) = A_{\rm SFR}\mathcal{F}_{\rm coll}(\boldsymbol{x},z) .
\end{equation}
This form is particularly helpful in the context of simulations, since it allows us to relate the local SFRD to the local density field.} According to the extended Press-Schechter (EPS) formalism~\cite{Press:1973iz,Bond:1990iw}, the collapsed fraction in Eq.~\eqref{eq:fcoll} can also be estimated based on the local density field smoothed over comoving scale $R=[3M_h/(4\pi \bar{\rho}_{m,0})]^{-3}$, which we denote as $\delta_{m,R}$. We have
\begin{equation}\label{eq:fcoll_eps}
    f_{{\rm coll},R}^{\rm EPS}(\boldsymbol{x},z,M_{\rm min}) = {\rm erfc}\left[\frac{\delta_c-\delta_{m,R}(\boldsymbol{x},z)}{\sqrt{\sigma^2(M_{\rm min},z)-\sigma^2(M_h,z)}}\right],
\end{equation}
where $\delta_c=1.686$ is the critical density for collapse, and the smoothed variance of the matter field is
\begin{equation}\label{eq:sigma}
    \sigma^2(M_h,z) = \int_0^\infty\frac{dk}{k}P_m(k,z)W_R^2(k),
\end{equation}
and we use a real-space spherical top-hat window function $W_R(k)=3[\sin(kR)-(kR)\cos(kR)]/(kR)^3$. 

In the original EPS formulation, the smoothed overdensity $\delta_{m,R}(\boldsymbol{x},z)$ is linearly evolved; we do the same in our \texttt{21cmFirstCLASS} simulation.
However, we note that using a non-linear evolution through 2nd-order Lagrangian perturbation theory~\cite{Friedman:1978ApJ,Buchert:1989} would be justified, since it seems to provide better agreement with radiative transfer simulations in the epoch of reionization~\cite{Flitter:2024eay}.

Combining Eqs.~\eqref{eq:fcoll}, \eqref{eq:hmf_to_fcoll}, and \eqref{eq:fcoll_eps}, we can estimate $\mathcal{F}_{\rm coll}^{\rm EPS}(\boldsymbol{x},z)$ and, if needed, scale this factor to any other HMF formalism through~\cite{Barkana:2003qk,Barkana:2007xj,Mesinger:2010ne}
\begin{equation}\label{eq:Fcoll}
    \mathcal{F}_{\rm coll}(\boldsymbol{x},z)=\mathcal{F}_{\rm coll}(z)\frac{\mathcal{F}_{\rm coll}^{\rm EPS}(\boldsymbol{x},z)}{\langle\mathcal{F}_{\rm coll}^{\rm EPS}(\boldsymbol{x},z)\rangle_{\boldsymbol{x}}}\,,
\end{equation}
where $\langle\mathcal{F}_{\rm coll}^{\rm EPS}(\boldsymbol{x},z)\rangle_{\boldsymbol{x}}$ is the spatial mean in the EPS case and $\mathcal{F}_{\rm coll}(z) \propto \int dM_h d\bar{n}/dM_h(z)M_h$ is proportional to the mean HMF. {The previous equation allows us to retain the density-dependent collapsed fraction from EPS (Eq.~\eqref{eq:fcoll_eps}) to model the local halo mass function, $dn(\boldsymbol{x},z)/dM_h$, while benefiting from the improved accuracy of other halo mass functions, which yield a different background-averaged value $dn(z)/dM_h$ compared to the original EPS prediction.} 
For our analysis in Sec.~\ref{sec:forecast}, we rely on the Sheth, Mo, and Tormen HMF~\cite{Sheth:1999mn,Sheth:1999su,Sheth:2001dp}
\begin{equation}
    \frac{d\bar{n}}{dM_h}(z) = -\frac{\Omega_m\rho_c}{M_h}\frac{d\log\sigma}{dM_h}\sqrt{2\pi}A_{\rm ST}(1+\hat{\nu}^{-2p_{\rm ST}})\hat{\nu}e^{-\hat{\nu}^2/2},
\end{equation}
where $\rho_c=3H_0^2 / (8\pi G)$ is the critical density, $A_{\rm ST}=0.353$, $p_{\rm ST}=0.175$, and $\hat{\nu}=\sqrt{0.73}\,\delta_c/\sigma$.

The spatial fluctuations in $T_{21}$ during cosmic dawn are mainly driven by inhomogeneities in the distribution of dark matter halos, where the astrophysical sources form.\footnote{Reference~\cite{Nikolic:2024xxo} recently showed that stochasticity in the astrophysical model has a non-negligible effect on the galaxy luminosity function at cosmic dawn. We expect $\Delta_{21}^2(k,z)$ to be similarly affected~\cite{Davies:2025}.} Therefore, $\Delta_{21}^2(k,z)$ is inevitably related to the HMF and hence to the underlying $P_m(k,z)$. Therefore, any deviation from $\Lambda$CDM, such as the ones caused by self-interacting neutrinos, should propagate to the 21-cm signal via the HMF.

\begin{figure}[t!]
    \centering
    \includegraphics[width=\columnwidth]{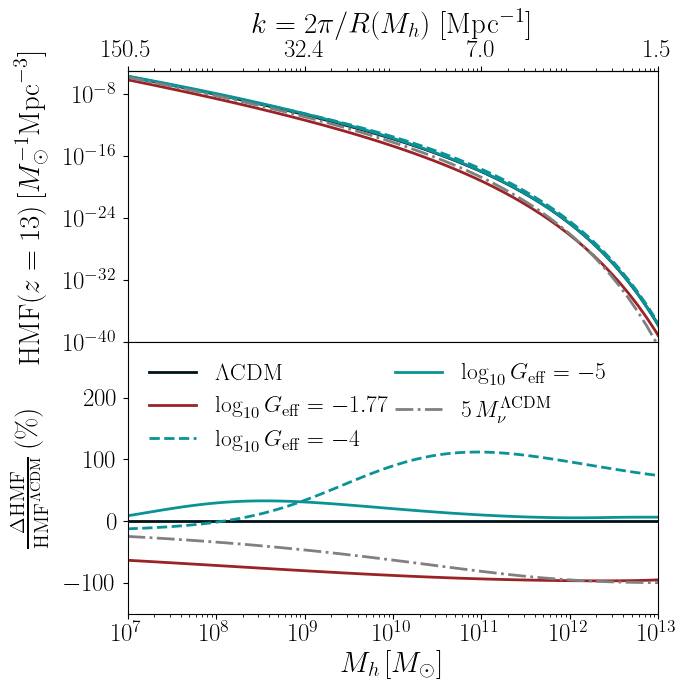}
    \caption{Halo mass function (top) and its residual with respect to $\Lambda$CDM (bottom). The legend is the same as in Fig.~\ref{fig:Cl}. Compared with Fig.~\ref{fig:Pm}, the effect each model has on the linear matter power spectrum at different scales propagates to variations in the abundance of halos with different $M_h$. {The boost the matter power spectrum experiences at large scales in the  $\Lambda$CDM+SI$\nu$ model is not represented here, since it increases the abundance of really massive halos, $M_h>10^{14}M_\odot$, which are extremely rare at the redshifts of interest for this work.}}
    \label{fig:pm_to_hmf}
\end{figure}

In Fig.~\ref{fig:pm_to_hmf}, we compare the HMF at cosmic dawn for $\Lambda$CDM with the HMF for the neutrino models of interest for our work.
The window function $W_R(k)$ in Eq.~\eqref{eq:sigma} provides a $k$-dependent weight that also depends on $R(M_h)$, so variations of $P_m(k,z)$ over certain $k$ regions roughly translate to variations of the HMF over certain halo masses $M_h$.
Due to the different values of $n_s$ between the SI$\nu$ model and the fiducial $\Lambda$CDM model (see Table~\ref{tab:fiducial_cosmo}), the HMF is suppressed for halo masses $M_h \lesssim 10^{14}\,{M_\odot}$.
The small boost of $P_m(k,z)$ at $k\sim 0.1\,{\rm Mpc}^{-1}$ in Fig.~\ref{fig:Pm} increases the abundance of halos with $M_h\gtrsim 10^{14}\,{M_\odot}$, which are very rare at high $z$; the 21-cm signal is almost blind to this high-mass end.

In contrast, the shape of $P_m(k,z)$ for the MI$\nu$ model implies that the number of mini-halos ($M_h\lesssim 10^8\,{M_\odot}$) is suppressed, while the number of halos with $M_h\gtrsim 10^9\,M_\odot$ is enhanced.
This result is particularly interesting for 21-cm at cosmic dawn, since galaxies in these halos drive the evolution of H in the IGM.
For the mI$\nu$ model, the smaller value of $G_{\rm eff}$ diminishes these features and shifts them to smaller masses, such that only halos of mass $M_h\in[10^8,10^{10}]\,{M_\odot}$ experience boost in the HMF. 

Finally, analogous to our discussion of $P_m(k,z)$ in Sec.~\ref{sec:nus_pk}, increasing $M_\nu$ has the opposite effect from the mI$\nu$ and MI$\nu$ models over a certain range of scales.
Additional neutrino mass suppresses the HMF for all masses above a certain threshold.
It is only at very small scales that the increased-$M_\nu$, mI$\nu$, and MI$\nu$ models all suppress the matter power spectrum, which translates into the suppression of very low-mass halos in Fig.~\ref{fig:pm_to_hmf}.

We now discuss the interpretation of the cosmic dawn signals $T_{21}(z)$ and $\Delta_{21}^2(k,z)$ in Figs.~\ref{fig:global_signal} and~\ref{fig:T21_pk_CD}, respectively, in light of our HMF results.
The change induced in the HMF by self-interacting neutrinos implies that the number of astrophysical sources is also varied, thus advancing or delaying the onset of cosmic dawn.

For example, the increased-$M_\nu$ and mI$\nu$ models shift $T_{21}(z)$ in different directions.
More massive neutrinos suppress small scales, reducing the abundance of halos in the small-to-intermediate mass range. Therefore, it takes more time to heat up the gas, and the difference between $T_s$ and the CMB temperature remains large down to a lower redshift, explaining why $T_{21}(z)$ and the amplitude of $\Delta_{21}^2(k,z)$ [shown at $z=13$ in Fig.~\ref{fig:T21_pk_CD}] are larger as compared to $\Lambda$CDM.
Note that the SI$\nu$ model suppresses the abundance of small-to-intermediate mass halos even more than the increased-$M_\nu$ model, further delaying cosmic dawn.
The mI$\nu$ model presents the exact opposite effect through its enhancement of power on scales relevant for the 21-cm signal, thereby advancing cosmic dawn.

It is important to note that the features in the matter power spectrum in Fig.~\ref{fig:Pm} propagate to features in the 21-cm power spectrum in Fig.~\ref{fig:T21_pk_CD} but at different values of $k$.
Small-scale deviations of the matter power spectrum appear at slightly larger scales in the HMF, which correspond to halos that host galaxies whose radiation fields impact the IGM and thus the 21-cm power spectrum at still larger scales.
Despite the complications it introduces in the modeling, the SFR is the key to move the indirect effects of the self-interacting neutrino models---and of small-scale physics in general---into regimes that can be actually probed by 21-cm and would be otherwise inaccessible by more direct observations of small-scale structure.

\vspace*{-.3cm}
\subsubsection{Astrophysical model}\label{sec:astro}

\begin{table}[t!]
    \centering
    \begin{tabular}{|cc|cc|}
    \hline
    \multicolumn{2}{|c|}{popII} & \multicolumn{2}{|c|}{popIII}\\
    \hline
    $\log_{10} f_{*,10}$ & -1.25 & $\log_{10} f_{*,7}$ & -2.5\\
    $\alpha_{*,10}$ & 0.5 & $\alpha_{*,7}$ & 0\\
    $\log_{10} f_{\rm esc,10}$ & 1.35& $\log_{10} f_{\rm esc,7}$ & -1.35\\
    $\alpha_{\rm esc,10}$ & -0.3 &$\alpha_{\rm esc,7}$ & -0.3 \\ 
    $\log_{10}(L_{X,10}/{\rm SFR})$ &  40.5 &  $\log_{10}(L_{X,7}/{\rm SFR})$ &  40.5\\
    \hline
    \end{tabular}
    \caption{Fiducial values of the astrophysical parameters adopted in the {\tt 21cmFirstCLASS} simulations for our analysis. For each type of source (popII or popIII stars), we provide the star formation efficiency $f_{*}$ and corresponding power-law index $\alpha_{*}$, the escape fraction $f_{\rm esc}$ and corresponding power-law index $\alpha_{\rm esc}$, and the soft-band X-ray luminosity per unit star formation rate $L_{X,{(10,7)}}/{\rm SFR}$. All quantities have a 10 or 7 subscript to denote the pivot masses $M_p^{\rm II} = 10^{10}M_\odot$ and $M_p^{\rm III} = 10^{7}M_\odot$ for the popII and popIII stars, respectively.}
    \label{tab:fiducial_astro}
\end{table}

Finally, to compute the 21-cm signal at cosmic dawn, we need to model the SFR $\dot{M}_*(M_h,z)$ of the first galaxies. The SFR is then used to calculate the radiation fields that drive the evolution of $T_s$ through the SFRD in Eq.~\eqref{eq:sfrd}.

We distinguish between {\it atomic cooling galaxies} (ACGs) and {\it molecular cooling galaxies} (MCGs). 
ACGs are produced inside halos with typical masses $M_h\simeq [10^9,10^{11}]\,M_\odot$, while MCGs are associated with mini-halos and thus form at higher $z$. Inside ACGs and MCGs, star formation is driven by different cooling mechanisms, and they can be roughly associated with population II (popII) and population III (popIII) stars, respectively~\cite{Tegmark:1996yt,Abel:2001pr,Bromm:2003vv,Trenti:2010hs}. 

Based on Refs.~\cite{Park:2018ljd,Munoz:2021psm}, we approximate the average SFR for each population as the stellar mass produced over a typical timescale: 
\begin{equation}\label{eq:sfr}
\begin{aligned}
    \dot{M}_*^{\rm II,III} &=\frac{M_*^{\rm II,III}}{0.5H^{-1}(z)}\\
    &=\frac{f_*^{\rm II,III}M_h{\Omega_b}/{\Omega_c}\left({M_h}/{M_p^{\rm II, III}}\right)^{\alpha^{\rm II,III}_*}}{0.5H^{-1}(z)}.
\end{aligned}
\end{equation}
Here, we have assumed that the stellar mass can be simply estimated by scaling the baryon mass inside a halo, $M_h\Omega_b/\Omega_c$, by a mass-dependent efficiency factor, $f_*^{\rm II,III}\left({M_h}/{M_p^{\rm II, III}}\right)^{\alpha^{\rm II,III}_*}$. This is a common approximation in the context of 21-cm simulations; see, e.g., Refs.~\cite{Libanore:2023oxf,Munoz:2023kkg,Cruz:2023rmo} for alternatives.

The pivot masses are $M_p^{\rm II} = 10^{10}M_\odot$ and $M_p^{\rm III} = 10^{7}M_\odot$; the values of $f_*^{\rm II,III}$ and $\alpha_*^{\rm II,III}$ are different for ACGs and MCGs and are summarized in Table~\ref{tab:fiducial_astro}.
The difference between popII and popIII stars also determines a different turnover mass $M_{\rm turn}^{\rm II,III}$ below which star formation becomes inefficient. This mass enters the calculation of the duty cycle $f_{\rm duty}(M_h)$ in Eq.~\eqref{eq:sfrd}, which is computed differently in the two scenarios:
\begin{align}
f_{\rm duty}^{\rm II}(M_h) &= \exp\left(-\frac{M_{\rm turn}^{\rm II}}{M_h} \right),\\
f_{\rm duty}^{\rm III}(M_h) &= \exp\left(-\frac{M_h}{M_{\rm turn}^{\rm III}} \right) \exp\left(-\frac{M_{\rm atom}}{M_h}\right).  
\end{align} 
These two expressions allow for a smooth transition of $f_{\rm duty}(M_h)$ between the halo-mass regime related to MCGs and the one in which ACGs dominate. The transition occurs at $M_{\rm atom}=3.3\times 10^7M_\odot[(1+z)/21]^{-3/2}$~\cite{Munoz:2021psm}. As for the turnover mass, in the case of ACG galaxies, its value is determined by stellar feedback, such as photoheating and supernovae~\cite{Hui:1997dp,Okamoto:2008sn,Ocvirk:2018pqh,Katz:2019due,Qin:2020xyh}.
{Based on simulation results~\cite{Sobacchi:2014}, we define 
\begin{equation}
    M_{\rm turn}^{\rm II}={\rm max}(M_{\rm atom},M_{\rm crit}^{\rm ion}),
\end{equation}
where
\begin{equation}
\begin{aligned}
    \frac{M_{\rm crit}^{\rm ion}}{2.8\times 10^9M_\odot}&=\left(\frac{\Gamma_{\rm ion}}{10^{-12}{\rm s}^{-1}}\right)^{0.17}\left(\frac{10}{1+z}\right)^{2.1}\times\\
    &\quad\quad\times\left[1-\left(\frac{1+z}{1+z_{\rm ion}}\right)^2\right]^{2.5},
\end{aligned}
\end{equation}
i.e., the critical halo mass $M_{\rm crit}^{\rm ion}$ below which star formation is inefficient is computed based on the local ionizing background $\Gamma_{\rm ion}$ and the reionization redshift $z_{\rm ion}$. On the other hand,}
for MCGs it is necessary to account for photons in the Lyman-Werner band (LW, 11.2--13.6 eV)~\cite{Tegmark:1996yt,Abel:2001pr,Bromm:2003vv,Trenti:2010hs,Kulkarni:2020ovu} and the suppression due to the relative velocity between dark matter and baryons~\cite{Tseliakhovich:2010bj,Naoz:2011if,
Stacy:2010gg,Bovy:2012af,Schmidt:2016coo}, 
{so that the turn over masses becomes
\begin{align}
    M_{\rm turn}^{\rm III}&={\rm max}(M_{\rm mol},M_{\rm crit}^{\rm ion}),\\
M_{\rm mol}&=\frac{3.3\times 10^7M_\odot}{(1+z)^{3/2}}f_{\rm LW}(J_{21}) f_{v_{\rm cb}}(v_{\rm cb})
\end{align}
In the previous equation, the factors $f_{\rm LW}(J_{21})$ and $f_{v_{\rm cb}}$ describe the LW feedback and relative-velocity feedback respectively. We describe through standard parameteric functions, 
\begin{align}
    f_{\rm LW}(J_{21}) &= 1+A_{\rm LW}(J_{21})^{\beta_{\rm LW}},\\
    f_{v_{\rm cb}}(v_{\rm cb}) &= \left(1+A_{v_{\rm cb}}\frac{v_{\rm cb}}{v_{\rm rms}}\right)^{\beta_{v_{\rm cb}}}, 
\end{align}
where $J_{21}$ describes the intensity of the LW photons, $v_{\rm rms}$ the rms velocity and $\{A_{\rm LW},\beta_{\rm LW},A_{v_{\rm cb}},\beta_{v_{\rm cb}}\}$ are free parameters; see, e.g.,~Ref.~\cite{Munoz:2021psm} for further detail. }

The last piece of information needed for the astrophysical model is represented by the escape fraction of ionizing photons from galaxies into the IGM, $f_{\rm esc}^{\rm II,III}(M_h)$. Following previous literature~\cite{Park:2018ljd}, we parameterize it to be constant in redshift and a function of $M_h$: 
\begin{equation}
    f_{\rm esc}^{\rm II,III}(M_h) = f_{{\rm esc},M_p^{\rm II,III}}^{\rm II,III}\left(\frac{M_h}{M_p^{\rm II,III}}\right)^{\alpha_{\rm esc}^{\rm II,III}}.
\end{equation}
Table~\ref{tab:fiducial_astro} collects the fiducial values of the astrophysical parameters, based on standard choices in the literature (see, e.g.,~Refs.~\cite{Park:2018ljd,Munoz:2021psm}).

\begin{figure}[t!]
    \centering
    \includegraphics[width=\columnwidth]{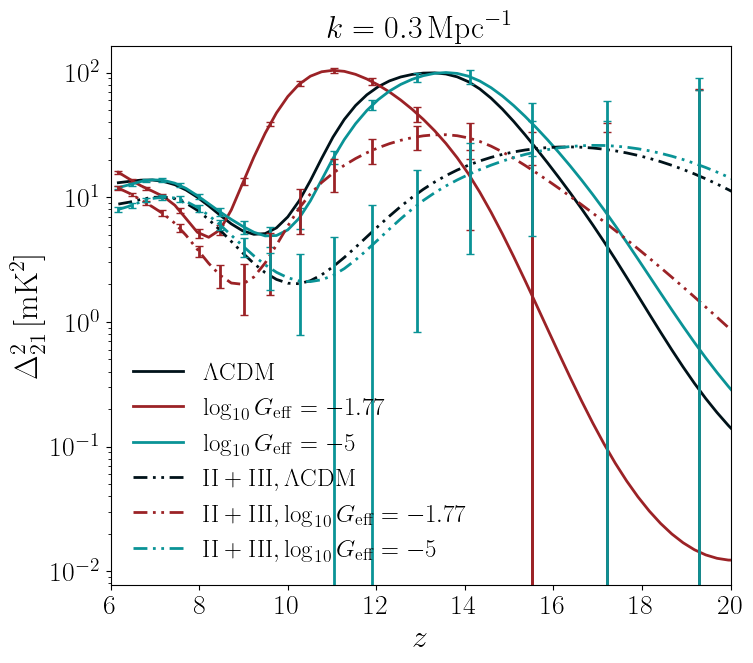}
    \caption{21-cm power spectrum during cosmic dawn for fixed $k$ and varying $z$. The color legend is the same as in Fig.~\ref{fig:Cl}. Solid lines represent our baseline (only popII stars), while the dashed lines include the contribution of popIII stars. The errorbars represent the sensitivity of the HERA detector we adopt in our analysis in Sec.~\ref{sec:forecast_detector}.}
    \label{fig:app_astro}
\end{figure}

Varying the astrophysical parameters affects, in a scale-independent way, the speed at which cosmic dawn proceeds, hence varying the position and depth of the trough in $T_{21}(z)$ in Fig.~\ref{fig:global_signal}. Similarly for $\Delta_{21}^2(k,z)$, the peaks rise at different $z$ with different heights in Fig.~\ref{fig:T21_pk_CD}.
As a further example, in Fig.~\ref{fig:app_astro}, we show $\Delta_{21}^2(k,z)$ for $\Lambda$CDM and for the SI$\nu$ and mI$\nu$ models, under a popII-only and a popII+popIII scenario. As expected, popIII stars anticipate the onset of cosmic dawn and the epoch of reionization; therefore, the peaks of $\Delta_{21}^2(k,z)$ are lower and shifted to higher $z$.


\begin{figure*}[t!]
    \centering
    \includegraphics[width=\columnwidth]{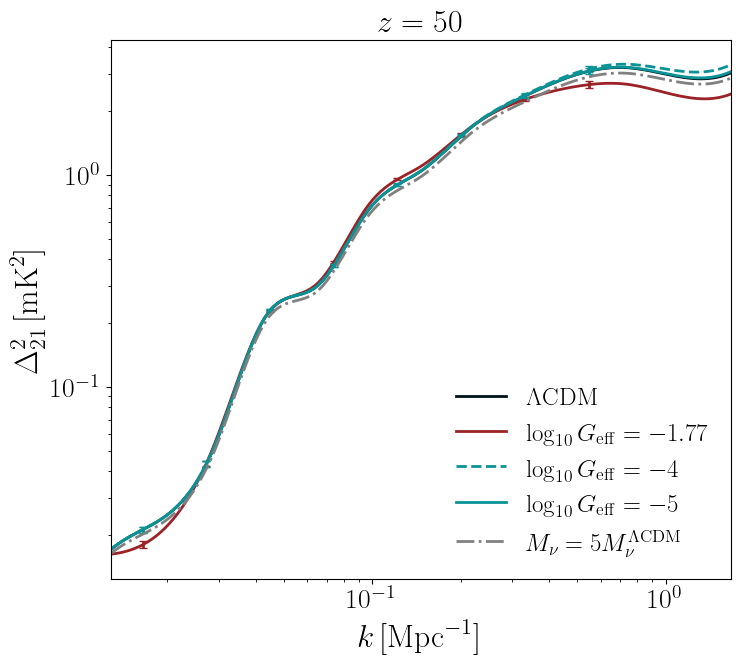}
\includegraphics[width=.95\columnwidth]{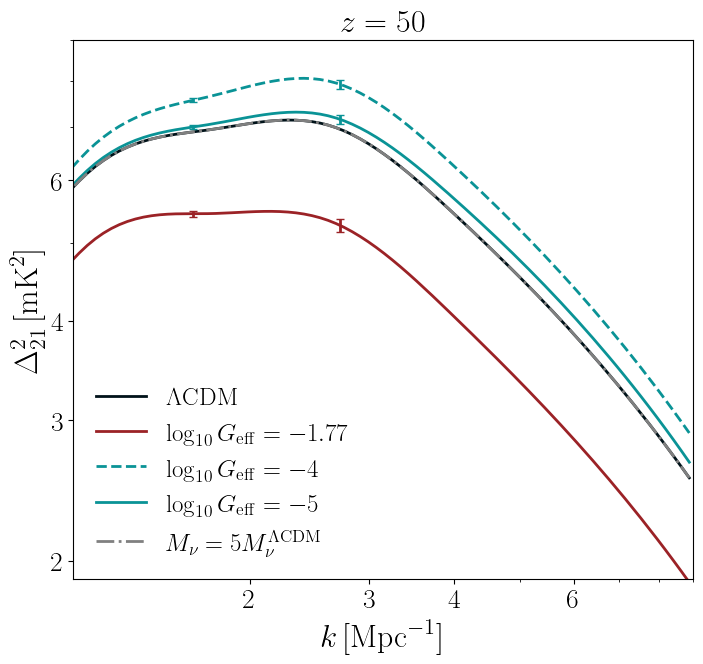}
    \caption{21-cm power spectrum during the dark ages for fixed $z$ and varying $k$. The legend is the same as in Fig.~\ref{fig:Cl}. The power spectrum for $k\lesssim 1\,{\rm Mpc^{-1}}$ (left) is extracted from the same simulation we generated for cosmic dawn. The power spectrum on smaller scales (right) is extracted from an additional \texttt{21cmFirstCLASS} simulation with a smaller box size and higher resolution; see Sec.~\ref{sec:results}. The errorbars show the sensitivity of the Lunar D (left) and LRA1 (right) detectors. The smallest $k$ is set by the size of the simulation box, and the largest by the detector resolution; their position depends on $\Delta\ln(k)$. Details in Sec.~\ref{sec:forecast_detector}.}
    \label{fig:T21_pk_DA}
\end{figure*}

\subsection{Dark ages}\label{sec:21cm_DA}

After recombination and before the formation of the first stars, the Universe is still sufficiently small that H collisional excitations are efficient: $T_s$ couples with $T_{\rm gas}$, and the 21-cm signal can be observed in absorption. The amplitude reaches its largest value around $z\sim 80$, but as the Universe continues expanding, the coupling efficiency and thus the amplitude of $T_{21}$ decrease until cosmic dawn begins.

The fluctuations $\delta T_{21}$ required to estimate the 21-cm power spectrum during the dark ages are given by~\cite{Ali-Haimoud:2013hpa}
\begin{equation}
\begin{aligned}
    \delta T_{21}(\boldsymbol{x},z) &= \frac{T_{21}(z)\delta_v+c_b(z)\delta_b+c_T(z)\delta_{T_k}}{T_{21}(z)}\\
    &\simeq \frac{dT_{21}}{d\delta_b}\delta_b(\boldsymbol{x},z)+T_{21}(z)\delta_v(\boldsymbol{x},z)
\end{aligned}
\end{equation}
where $\delta_b(\boldsymbol{x},z)$ is the baryon density contrast, $\delta_v(\boldsymbol{x},z)=-(1+z)\partial_rv/H(z)$ is the peculiar velocity fluctuation, and $\delta_{T_k}(\boldsymbol{x},z)$ is the spatial fluctuation of the gas kinetic temperature. In the second line, we neglect the fluctuations in the gas temperature.

Before recombination, baryons are coupled with CMB photons and therefore oscillate without collapsing; after they decouple, baryons begin to follow the dark matter distribution and can be treated as a biased tracer of the underlying dark matter field. However, it is important to account for some subtleties.

First, during dark ages, the coupling between baryons and dark matter is not yet complete, and the evolution of $\delta_b(\boldsymbol{k},z)$ must be estimated using the scale-dependent growth factor $\mathcal{D}_b(k,z)=\mathcal{T}_b(k,z)/\mathcal{T}_b(k,z=0)$, where $\mathcal{T}_b(k,z)$ is the baryon transfer function. Using this growth factor has a non-negligible impact on the computation of $\Delta_{21}^2(k,z)$, compared to the case in which the signal is computed using the total matter fluctuation $\delta_m(k,z)$ or under the approximation of scale-independent growth factor, $\delta_m(k,z)=\delta_0(k)D(z)$~\cite{Flitter:2023rzv, Flitter:2024eay}.

Moreover, the relative velocity between baryons and dark matter suppresses small-scale power in $P_m(k,z)$~\cite{Tseliakhovich:2010bj,Naoz:2011if,
Stacy:2010gg,Bovy:2012af,Schmidt:2016coo,Ali-Haimoud:2013hpa,Mu_oz_2019}. This suppression propagates to all scales in $\Delta_{21}^2(k,z)$, due to the non-linear relation between $\delta T_{21}$ and $\delta_b$ and due to a large-scale modulation that velocities induce on $\delta_b$. The suppression effect is larger at cosmic dawn than during the dark ages.

The 21-cm power spectrum during the dark ages can be theoretically estimated by solving the Boltzmann equations. Its shape fully depends on the underlying cosmological model, particularly on the form of $P_m(k,z)$.
However, for consistency, we extract $\Delta_{21}^2(k,z)$ from the same {\tt 21cmFirstClass} runs realized for the cosmic dawn.

As an example, Fig.~\ref{fig:T21_pk_DA} shows $\Delta_{21}^2(k,z)$ at $z=50$ for $\Lambda$CDM and the neutrino models of interest.
Note that the deviations of $\Delta_{21}^2(k,z)$ induced by the neutrino models on the dark ages signal occur on scales $k$ more comparable (as compared to the cosmic dawn signal) to those in $P_m(k,z)$.
As shown in Fig.~\ref{fig:global_signal}, the self-interacting neutrino models do not affect $T_{21}(z)$ during the dark ages, since there is no mechanism to impact the evolution of the gas temperature. Variations on $\Delta_{21}^2(k,z)$ are directly sourced by the shape of $P_m(k,z)$ shown in Fig.~\ref{fig:Pm}: the SI$\nu$ power suppression is due to the smaller value of $n_s$ relative to $\Lambda$CDM, while the enhancements for MI$\nu$ and mI$\nu$ reflect the bumps in their corresponding $P_m(k,z)$.

To better capture the effects of MI$\nu$ and mI$\nu$ in the dark ages, we run a higher-resolution simulation, as discussed in Sec.~\ref{sec:forecast}.
The resulting 21-cm power spectrum is shown in the right panel of Fig.~\ref{fig:T21_pk_DA}.
Features in the MI$\nu$ and mI$\nu$ models are in principle distinguishable from $\Lambda$CDM for $k\gtrsim 1\,{\rm Mpc}^{-1}$, but detecting these scales would require detectors with very high angular resolution.


\section{Running forecasts}\label{sec:forecast}

To simulate the 21-cm signal during both the dark ages and cosmic dawn, we use \texttt{21cmFirstCLASS}. The simulation allows us to account for different initial cosmologies, for the effects of the baryon-dark matter relative velocity (suppression of the matter power spectrum and of the minimum halo mass required to form stars~\cite{Tseliakhovich:2010bj,Naoz:2011if,
Stacy:2010gg,Bovy:2012af,Schmidt:2016coo,Ali-Haimoud:2013hpa,Mu_oz_2019}; mode coupling during the dark ages instead is not included at this stage), for the scale-dependent growth factor and for the evolution of baryon- and dark-matter perturbations. 

We choose a box size of $L = 600\,{\rm Mpc}$ and a resolution $N_{\rm cell} = 256$, which provides a $k$-space resolution between $k_{\rm min}=2\pi/L_{\rm box}\simeq 0.01\,{\rm Mpc}^{-1}$ and $k_{\rm max}=2\pi N_{\rm cell}/L_{\rm box}\simeq 2.5 \,{\rm Mpc}^{-1}$. This setup allows us to simulate the signal in the sensitivity band with sample variance smaller than the detector noise (see Sec.~\ref{sec:forecast_detector}).

We run a second simulation with $L = 50\,{\rm Mpc}$ and $N_{\rm cell} = 128$ to improve the resolution for our dark ages analysis, accessing scales $0.15\,{\rm Mpc}^{-1}\lesssim k\lesssim 10\,{\rm Mpc}^{-1}$.

The initial conditions for the simulation are generated by our \texttt{nuCLASS} code, which incorporates the effects of neutrino self-interactions.
The transfer function and growth factor produced by \texttt{nuCLASS} are used in \texttt{21cmFirstCLASS} to evolve the simulation consistently with the input cosmology. Since self-interacting neutrinos do not interact with dark matter and baryons, the Boltzmann equations inside the simulation are unaffected.
Within \texttt{nuCLASS}, we consider three degenerate neutrino species and describe their properties in terms of the coupling strength $G_{\rm eff}$, the sum of their masses $M_\nu$, and the effective number of relativistic neutrino-like species $N_{\rm eff}$.

\subsubsection{Fiducial setup}

The fiducial cosmological parameters we use for our analyses are listed in Table~\ref{tab:fiducial_cosmo}.
In particular, we use an optical depth $\tau$ that is produced as an output by our 21-cm simulations; it is then used to compute the CMB power spectrum for our CMB analyses.

The fiducial astrophysical parameters we use for our cosmic dawn analyses are listed in Table~\ref{tab:fiducial_astro}.
We consider two scenarios: popII-only and popII+popIII. The popII-only scenario, in which the evolution of $T_s$ is driven by atomic-cooling galaxies, is used in the official HERA analysis pipeline~\cite{HERA:2021bsv,HERA:2021noe}; therefore, we consider it to be our baseline. Including popIII stars, which are sourced by molecular cooling galaxies inside mini-halos with $M_h \lesssim 10^8\,M_\odot$~\cite{Ciardi:2005gc,Kimm:2016kkj,Qin:2020xyh}, anticipates reionization, and $\tau$ shifts to $\sim 0.05$ in the SI$\nu$ model and to $\sim 0.07$ in the others.
The addition of popIII stars also increases the number of parameters, so the analysis is subject to larger uncertainties and is computationally more expensive. Nevertheless, the abundance of mini-halos strongly depends on the shape of $P_m(k,z)$ on very small scales, so we expect the inclusion of popIII stars to affect the constraining power of HERA data~\cite{Lazare:2023jkg}, particularly for the MI$\nu$ model.
Therefore, we also analyze a popII+popIII scenario.


\vspace*{-.3cm}
\subsubsection{Optical depth computation}\label{sec:forecast_tau}
\vspace*{-.15cm}

The optical depth $\tau$ is the line-of-sight integral of the mean electron density $\bar{n}_e(z)$ and is given by~\cite{Liu:2015txa,Shmueli:2023box}
\begin{equation}\label{eq:tau}
\begin{aligned}
    \tau = &\frac{\sigma_T\,3H_0^2\Omega_b}{8\pi G m_p}\left[1+\frac{Y_p^{\rm BBN}}{4}\left(\frac{m_{\rm He}}{m_{\rm H}}-1\right)\right]\times\\
    &\times\int_0^{z_{\rm CMB}} dz 
    \frac{c(1+z)^2\,\overline{x_{\rm HII}(1+\delta_b)}}{H_0\sqrt{\Omega_m(1+z)^3+\Omega_\Lambda}},
\end{aligned}
\end{equation}
where $\sigma_T$ is the Thompson cross-section, $m_{\rm p,H,He}$ are the proton mass and the atomic masses for H and Helium (He), respectively, and $Y_p^{\rm BBN}$ is the primordial He mass fraction; {the previous expression has been derived under the assumption that He is instantaneously reionized at $z=3$ and uncertainties in this process are negligible}. 
We can estimate $\tau$ and its uncertainty based on the other cosmological parameters and the density-weighted ionization fraction $\overline{x_{\rm HII}(1+\delta_b)}$~\cite{Liu:2015txa,Shmueli:2023box}.
This latter quantity is approximately $\sim 1$ for $z\lesssim 5$ when the Universe is fully ionized and $\sim 0$ for $z\gtrsim 35$ before reionization; its value during cosmic dawn and the epoch of reionization can instead be directly estimated from our \texttt{21cmFirstCLASS} simulations.
The resulting value of $\tau$ for each of our fiducial models in Table~\ref{tab:fiducial_cosmo} is consistent with the cosmological and astrophysical model in the simulation.
In our analysis in Sec.~\ref{sec:forecast_fisher}, the uncertainty on $\tau$ is inferred from the uncertainties on the other parameters. 


\subsection{Detector specifications}\label{sec:forecast_detector}

The sensitivity of detectors measuring the 21-cm signal or the CMB has been widely discussed in the literature. Here, we summarize the main ingredients needed for our forecasts.

\begin{figure}[t!]
    \centering
    \includegraphics[width=\columnwidth]{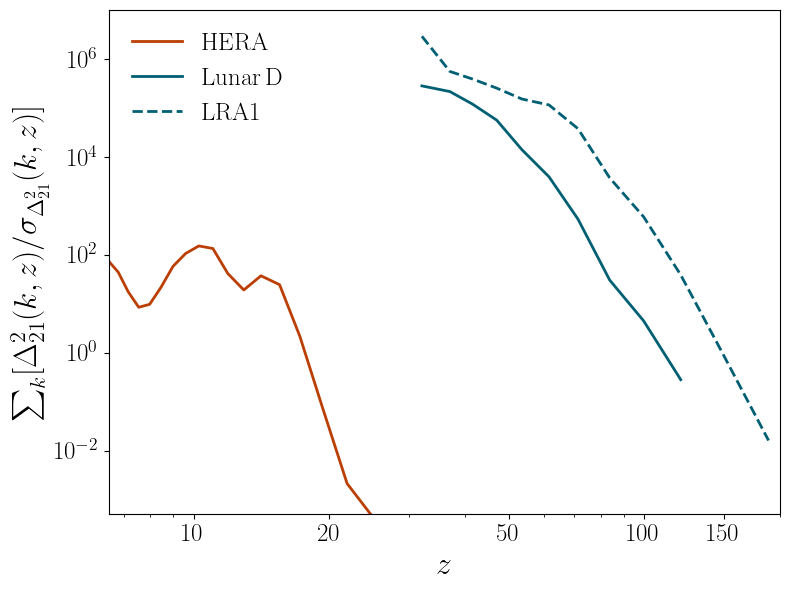}
    \vspace{-0.2in}
    \caption{Cumulative SNR as a function of $z$ for the 21-cm experiments considered in this work. HERA probes the era of cosmic dawn, while Lunar D and LRA1 are configurations to probe the dark ages.}
    \label{fig:noise_21cm}
\end{figure}

\subsubsection{Cosmic dawn: HERA}
\vspace*{-.2cm}
To explore the constraining power of the 21-cm signal during cosmic dawn, we consider the design sensitivity of HERA~\cite{DeBoer:2016tnn}, which is expected to be achieved in the reasonably near future.
It is composed of $\sim 330$ packed antennas, distributed in a hexagonal shape with 14\,m diameter dishes. 
The observational frequency band ranges between 50 and 250\,MHz, with a resolution 8\,MHz and 82 channels probing each bandwidth.
The 21-cm signal is divided into $\sim 25$ bins between $z\sim[5,27]$. The bins are equally spaced in frequency, while their width varies in redshift.
We assume $t_o\simeq 3000\,$hr integration time (540 days). The detector properties are summarized in Table~\ref{tab:detectors}.

We estimate the sensitivity $\sigma_{\Delta_{21}^2,{\rm CD}}$ using the public code \texttt{21cmSense}\footnote{\url{https://github.com/jpober/21cmSense}}~\cite{Pober:2012zz,Pober:2013jna,Murray:2024the}.
The code allows us to account for cosmic variance, instrumental noise, and foreground reduction. In particular, we assume the so-called {\it moderate scenario}~\cite{Pober:2013jna}, in which the foreground-contaminated wedge on the line-of-sight modes $k_\parallel$ extends 0.1\,$h$Mpc$^{-1}$ beyond the detector horizon~\cite{Parsons:2012qh}.

Figure~\ref{fig:noise_21cm} shows the redshift-dependent cumulative signal-to-noise ratio 
\begin{equation}
    {\rm SNR}(z) = \sum_{k\in k_{\rm HERA}}\frac{\Delta_{21}^2(k,z)}{\sigma_{\Delta_{21}^2,{\rm CD}}(k,z)}    
\end{equation}
for HERA, obtained using \texttt{21cmSense} and summing over the scale range $k_{\rm HERA}\in[0.05,1.5]\,{\rm Mpc}^{-1}$ allowed by the instrument resolution.

\subsubsection{Dark ages: Lunar arrays}

Finally, we explore the prospects of measuring the 21-cm signal deep in the dark ages. The SKAO or any other ground-based detector can only slightly extend the low-frequency range observed by HERA because of atmospheric effects. Hence, a new observational window can be opened by constructing an array in a location with no atmosphere, such as the Lunar far side, which would also use the Moon to shield radio interference from Earth (e.g.,~Refs.~\cite{Burns:2021pkx,Chen:2020lok,Chen:2024tvn,Saliwanchik:2024nrp}). 

We provide an order-of-magnitude estimate of the noise that such detectors could reach. Following Refs.~\cite{Mellema:2012ht,Mondal:2015oga,Mondal:2023xjx} (see also Refs.~\cite{Lidz:2007az,Mondal:2016hmf,McQuinn:2005hk} for further details on the 21-cm noise computation), we model the noise $\sigma_{\Delta_{21}^2,\rm DA}(k,z)$ as the sum between the thermal noise
\begin{equation}\label{eq:noise_DA_1}
    \Delta_N^2(k,z) = \frac{2}{\pi}\sqrt{\frac{k^3V_z}{\Delta\ln k}}
    \frac{T_{\rm sys}^2}{\delta\nu \,t_{o}}\frac{1}{N_{\rm ant}}
\end{equation}
and the sample variance
\begin{equation}\label{eq:noise_DA_2}
   \Delta_{CV}^2(k,z) = \frac{2\pi \Delta^2_{21}(k,z)}{\sqrt{V_z k^3\Delta\ln k}},
\end{equation}
which we set equal to the signal mean in the Poisson approximation. 
The observed volume is $V_z= 4\pi f_{\rm sky}\chi^2(z)\Delta {\chi(z)}$, where $\chi(z)$ is the comoving distance to the center of each redshift bin $z_i$, and $\Delta\chi = \chi(z_{\rm max}^i)-\chi(z_{\rm min}^i)$ is its width. We set the observed sky fraction to $f_{\rm sky} = 0.75$ and the system temperature to the sky temperature $T_{\rm sys} \sim T_{\rm sky} = 180\,{\rm K}(\nu/180\,{\rm MHz})^{-2.6}$, neglecting the contribution of the instrument. 

In our dark ages forecast, we consider two different detector configurations: Lunar D (based on the D scenario in the Supplementary Material of Ref.~\cite{Mondal:2023xjx}) and LRA1 (based on the lunar radio array in Refs.~\cite{Bernal:2017nec,  Vanzan:2023gui,deKruijf:2024voc,Vanetti:2025aca}).
For Lunar D, we use 10 frequency bins between $10\,{\rm MHz}$ and $45\,{\rm MHz}$, with frequency resolution $\delta\nu = 5\,{\rm MHz}$. These choices correspond to $z \sim [30,150]$, with bins between $z_{\rm min,max}^i = -1+\nu_{21}/(\nu_{\rm obs}^i\pm \delta\nu/2)$. For LRA1, we use 40\,bins between $6.3\,{\rm MHz}$ and $46.3\,{\rm MHz}$ (corresponding to $z\sim [30,220]$), with frequency resolution $\delta\nu = 1\,{\rm MHz}$.

For both configurations, we set the total observation time to $t_o = 10^4\,{\rm hr}$.
We estimate the number of dipole-like antennas to be~\cite{Zaldarriaga:2003du}
\begin{equation}
    N_{\rm ant} = \frac{f_{\rm cover}\pi{D_{\rm base}^2}/{4}}{A_{\rm eff}},
\end{equation}
where $D_{\rm base}$ is the maximum baseline of the experiment, $f_{\rm cover}$ is the fraction of interferometer area covered by antennas, and $A_{\rm eff}$ is the effective collecting area of the dipoles.
We estimate $ A_{\rm eff} =G{(21\,{\rm cm})^2(1+50)^2}/{4\pi}$ with $G=1.64$. Therefore, in our analyses we consider $\{D_{\rm base}, f_{\rm cover},\,N_{\rm ant}\} = \{20\,{\rm km},0.2,4$$\times$$10^6\}$ for Lunar D and $\{100\,{\rm km},0.2,1$$\times$$10^8\}$ for LRA1. All detector properties are summarized in Table~\ref{tab:detectors}.

\begin{table}[t!]
    \hspace*{-.15cm}
    \begin{tabular}{|c|ccccc|}
    \hline
         & $\Delta\nu$\,[MHz] & $\delta\nu$\,[MHz] & $N_{\rm ant}$& $t_o$\,[hr]& Ref. \\
    \hline
        HERA  & [50,\,250] & 8 & 330 dishes & 3000 & \cite{Pober:2012zz,Pober:2013jna}\\
    \hline
    Lunar\,D & [10,\,45]  & 5 & $4$$\times$$10^6$\,dipoles & $10^4$ & \cite{Mondal:2023xjx}\\
    LRA1 & [6.3,\,46.3] & 1 & 1$\times$$10^8$\,dipoles & $10^4$ & \cite{Vanzan:2023gui,deKruijf:2024voc,Vanetti:2025aca} \\
    \hline
    \end{tabular}
    \caption{Specifications for the 21-cm detectors used in our analyses. We list the observed frequency range $\Delta\nu$, frequency resolution $\delta\nu$, type and number of antennas $N_{\rm ant}$, and observation time $t_o$. }
    \label{tab:detectors}
\end{table}

Finally, we set $\Delta \ln k = 0.5$ and define $k_{\rm min} = 2\pi/L_{\rm box}$ and $k_{\rm max}$ as the maximum between the scale probed by the resolution of the simulation, $2\pi N_{\rm cell}/L_{\rm box}$, and the resolution of the detector, $2\pi D_{\rm base}/[21\,{\rm cm}(1+z)\chi(z)]]$. For example, at $z = 30$, we have $k_{\rm max} \simeq 1.5\,{\rm Mpc}^{-1}$ for Lunar D and $k_{\rm max} \simeq 8\,{\rm Mpc}^{-1}$ for LRA1; therefore, for Lunar D, we derive our results using the same \texttt{21cmFirstCLASS} simulation adopted for cosmic dawn, while for LRA1 we add the information on scales $k> 1.5\,{\rm Mpc}^{-1}$ from the smaller simulation introduced in at beginning of Sec.~\ref{sec:forecast}.

Figure~\ref{fig:noise_21cm} compares the performance of the two Lunar setups with HERA. Although the Lunar detector is more futuristic, we see that it would allow us to probe a completely different regime of cosmic history, where the cosmological signal is pristine: it is in the linear regime with no astrophysical contamination.

\subsubsection{CMB-S4}
For the CMB analysis, we adopt the latest specifications of the wide survey proposed for CMB-S4. 
We use noise curves from the {\it CMB-S4: Dark Radiation Anisotropy Flowdown Team (DRAFT) tool}\,\footnote{\url{https://github.com/sriniraghunathan/DRAFT/tree/master}}, for the {\it S4-wide Chilean LAT configuration}\footnote{\url{https://github.com/sriniraghunathan/DRAFT/tree/master/products/202310xx_PBDR_config/s4wide_202310xx_pbdr_config}}, updated on October 2023~\cite{raghunathan2023crossinternallinearcombinationapproach,Green:2016cjr,Hotinli:2021umk,SPT:2020psp}. Here, the observed sky fraction is $f_{\rm sky} = 0.6$, and the properties of the detector are combined separately for TT and EE to get the noise power spectra
\begin{equation}
    \begin{aligned}N_\ell^{T,E} = &\Delta_{T,E}^2[{\rm rad}]\exp\left(\ell(\ell+1)\frac{\theta_{\rm FWHM}[{\rm rad}]}{\sqrt{8\log(2)}}\right)\\
    &\times\left[1+(\ell\ell_{\rm knee}^{T,E})^{\alpha_{T,E}}\right].
\end{aligned}
\end{equation}
We use the 93\,GHz channel, for which the angular resolution is $\theta_{\rm FWHM}= 2.2\,{\rm arcmin}$, and the instrumental noises are $\Delta_T=2\,{\rm \mu K/arcmin}$ and $\Delta_E=2.68\,{\rm \mu K/arcmin}$. Additionally, $[\ell_{\rm knee}^T,\alpha_T] = [1932,3.5]$ and $[\ell_{\rm knee}^E,\alpha_E] = [700,1,4]$. We restrict our analysis to multipoles $\ell \in [30,3000]$.

We could improve sensitivity, for example, by implementing a cross-internal linear combination approach~\cite{raghunathan2023crossinternallinearcombinationapproach} or using a forecast-loop pipeline~\cite{CMB-S4:2020lpa}.
These approaches can further improve the constraining power on the small scales by reducing the contamination of astrophysical signals; their implementation, however, is beyond the scope of this work.


\subsection{Fisher matrix formalism}\label{sec:forecast_fisher}

We use the Fisher matrix formalism to perform our forecasts (see, e.g.,~Ref.~\cite{Verde:2009tu} for review). Our baseline cosmological parameters and their fiducial values are listed in Table~\ref{tab:fiducial_cosmo}. We have the six vanilla $\Lambda$CDM cosmological parameters and the three parameters that characterize the neutrino properties: $G_{\rm eff}$, $M_\nu$, and $N_{\rm eff}$.
We also include the full set of astrophysical parameters listed in Table~\ref{tab:fiducial_astro}.

For each self-interacting neutrino model (SI$_\nu$, MI$_\nu$, mI$_\nu$) and each astrophysical scenario (popII, popII+popIII), we estimate the numerical derivatives of $\Delta_{21}^2(k,z)$ with respect to all parameters using different realizations of \texttt{21cmFirstCLASS}.
Since the optical depth $\tau$ does not influence either the shape or the amplitude of the 21-cm power spectrum, we have $d\Delta_{21}^2/d\tau=0$, indicating that $\tau$ cannot be constrained through the 21-cm signal alone. However, as discussed in Sec.~\ref{sec:forecast_tau}, the value of $\tau$ at cosmic dawn directly depends on the ionization fraction $x_{\rm HII}$; hence, we estimate it from the output of our \texttt{21cmFirstCLASS} simulations.

Using the $\sigma_{\Delta_{21}^2,({\rm CD,DA})}(k,z)$ sensitivities estimated by \texttt{21cmSense} for cosmic dawn or from Eqs.~\eqref{eq:noise_DA_1} and \eqref{eq:noise_DA_2} for the dark ages, the Fisher matrices for HERA and Lunar D are
\begin{equation}\label{eq:fisher_21cm}
    F^{({\rm CD,DA})}_{21,\alpha\beta} = \sum_{z,k} \frac{1}{\sigma^2_{\Delta^2_{21},({\rm CD,DA})}(k,z)}\frac{d\Delta^2_{21}(k,z)}{d\theta_\alpha}\frac{d\Delta^2_{21}(k,z)}{d\theta_\beta}.
\end{equation}
Finally, forecasts for LRA1 are obtained by summing two Fisher matrices: one estimated from the same simulation box as the other detectors and another from the small, high-resolution box. To avoid double counting when summing over $k$, we use the larger box up to $k=1\,{\rm Mpc}^{-1}$ and the smaller box to account for larger $k$.
The $1\sigma$ uncertainty for each cosmological or astrophysical parameter $\theta_i$ is obtained from the diagonal elements of the inverse Fisher matrices:
\begin{equation}
    \sigma_{\theta_{i},(\rm CD,DA)}=\sqrt{\left(F^{(\rm CD,DA)}_{21}\right)^{-1}_{ii}}.
\end{equation}

We construct the CMB-S4 Fisher matrix for each model of interest by running \texttt{nuCLASS} to get the covariance and the derivatives of $C_\ell^{TT,TE,EE}$ with respect to the cosmological parameters. We choose the same step sizes for the computation of the 21-cm and CMB numerical derivatives. In this case, the Fisher matrix is
\begin{equation}
    F_{\rm CMB,\alpha\beta} = f_{\rm sky}\sum_\ell \frac{2\ell+1}{2}{\rm Tr}\left[\bold{C}_\ell^{-1}\frac{d\bold{C}_\ell}{d\theta_\alpha}\bold{C}_\ell^{-1}\frac{d\bold{C}_\ell}{d\theta_\beta}\right]
\end{equation}
where the $\bold{C}_\ell$ matrix is defined by the signal and noise contributions in the TT, TE, and EE channels:
\begin{equation}
    \bold{C}_\ell = \begin{pmatrix}
        C_\ell^{TT} + N_\ell^{TT} & C_\ell^{TE}\\
        C_\ell^{TE} & {C}_\ell^{EE}+ N_\ell^{EE}\\
    \end{pmatrix}.
\end{equation}
Again, we obtain the $1\sigma$ uncertainty from the diagonal elements of the inverted Fisher matrix
\begin{equation}
    \sigma_{\theta_i,\rm CMB}=\sqrt{\left(F_{{\rm CMB}}\right)^{-1}_{ii}}.
\end{equation}
Since we are using a single channel and a restricted range $\ell\in[30,3000]$, our forecasts are conservative. 

Finally, we combine the 21-cm and CMB Fisher matrices to help break parameter degeneracies. 
For the dark ages, we use the matrix $F_{\rm tot}^{DA} = F_{\rm CMB} + F_{21}^{\rm DA}$.
For cosmic dawn, we must account for the propagation of uncertainties in the modeling of $\tau$ in Sec.~\ref{sec:forecast_tau}.
Following Refs.~\cite{Liu:2015txa,Shmueli:2023box}, we first compute $F_{\rm tot}^{\rm CD} = F_{\rm CMB} + F_{21}^{\rm CD}$ (including all astrophysical and cosmological parameters) and then define
\begin{equation}\label{eq:fisher_total}
    \tilde{F}_{{\rm tot},ij}^{\rm CD} = F_{{\rm tot},ij}^{\rm CD}+a_i F_{{\rm tot},j\tau}^{\rm CD}+ a_j F_{{\rm tot},i\tau}^{\rm CD} + a_ia_jF_{{\rm tot},\tau\tau}^{\rm CD}.
\end{equation}
In $\tilde{F}^{\rm CD}_{\rm tot}$ there are no rows and columns related to $\tau$, and for each parameter $\theta_i$, the factor $a_i = d\tau/d\theta_i$ is estimated from the different realizations of the \texttt{21cmFirstCLASS} simulation. The uncertainty in the optical depth is then computed by drawing random samples of all the other parameters $\theta_i$ from Gaussian distributions centered around their fiducial values with a standard deviation $\sigma_{{\rm tot},\theta_i}=\sqrt{(\tilde{F}^{\rm CD}_{{\rm tot}})^{-1}_{ii}}$. Thus, for each sample, we estimate the optical depth as $\tilde{\tau}=\sum_{\theta_i}a_i\theta_i$, and we set the uncertainty $\sigma_{\rm tot,\tau}$ to be equal to the spread between the different values of $\tilde{\tau}$.


\vspace*{-.3cm}
\section{Results}\label{sec:results}

\begin{table}[t!]
    \centering
    \begin{tabular}{|cc|c|c|c|}
\hline
    & & {$\quad$SI$_\nu$$\quad$} & {$\quad$MI$_\nu$$\quad$} & {$\quad$mI$_\nu$$\quad$} \\
\hline
    \multicolumn{2}{|c|}{$\log_{10}G_{\rm eff}$ fiducial value} &  -1.77 & -4 & -5 \\ 
    \hline
    \hline
    \multicolumn{2}{|c|}{CMB-S4} & 14\% & 66\% & 534\% \\ 
    \hline
    \hline
    \multirow{2}{*}{HERA (cosmic dawn)} & II & 32\% & 16\% & 112\% \\ 
    & {II+III} & {72\%} & {176\%} & {78\%}\\
    \hline
    \multirow{2}{*}{HERA + CMB-S4} & II & 8\% & 11\% & 14\% \\ 
    & {II+III} & {13\%} & {31\%} & {24\%}\\
    \hline
    \hline
    \multicolumn{2}{|c|}{Lunar D (dark ages)} & 14\% & 42\% & 240\% \\ 
    \multicolumn{2}{|c|}{LRA1 (dark ages)}& 1\% & 2\% & 5\% \\ 
    \hline
    \multicolumn{2}{|c|}{Lunar D + CMB-S4} & 5\% & 19\% & 123\% \\ 
    \multicolumn{2}{|c|}{LRA1 + CMB-S4} & 1\% & 1\% & 4\% \\ 
\hline
    \end{tabular}
    \caption{Summary of the main results of our paper. We provide forecasts for the 2$\sigma$ uncertainties on the (log of the) neutrino self-coupling strength $\log_{10}(G_{\rm eff}/{\rm MeV}^{-2})$, for strong (SI$\nu$), moderate (MI$\nu$), and mild (mI$\nu$) self-interaction models; see Table~\ref{tab:fiducial_cosmo}. The MI$\nu$ and mI$\nu$ models are at the cusp or beyond the reach, respectively, of current CMB experiments. {The results for the cosmic dawn analyses use either the popII or the popII+popIII astrophysical models in Table~\ref{tab:fiducial_astro}.}}
    \label{tab:results}
\end{table}

\begin{figure*}[ht!]
    \centering
    \includegraphics[width=0.68\columnwidth]{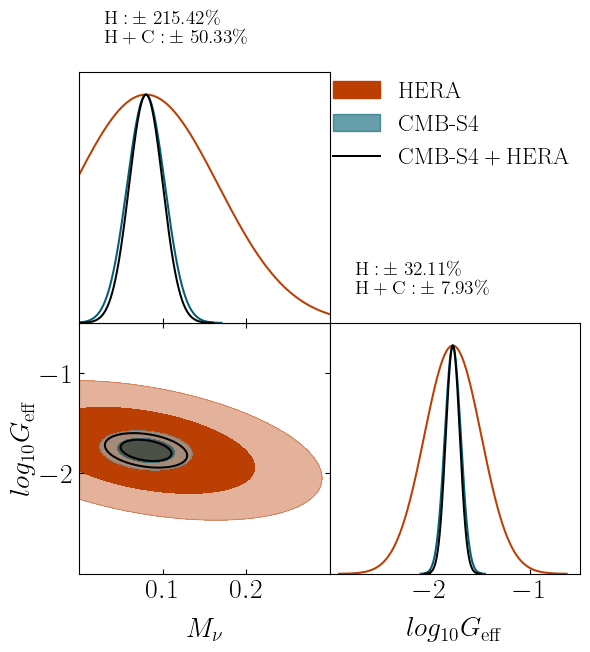}
    \includegraphics[width=0.68\columnwidth]{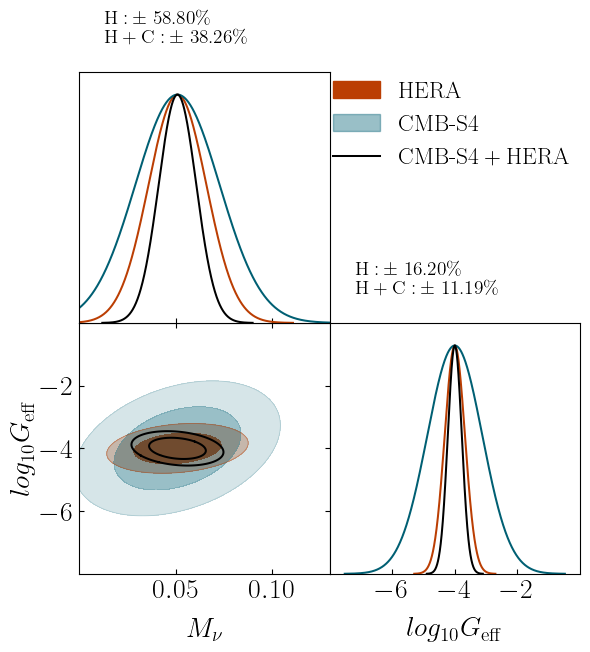}
    \includegraphics[width=0.68\columnwidth]{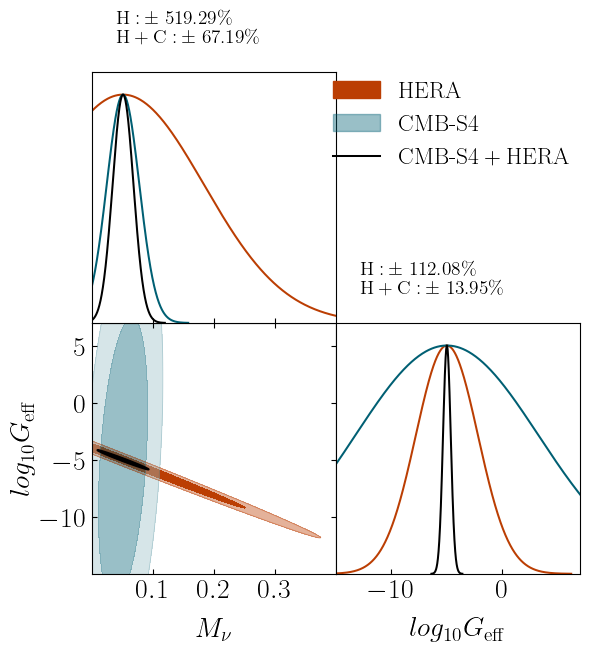}
    \vspace{-0.1in}
    \caption{Summary plot of the main results of our paper. We show forecasts for the constraining power of $\Delta^2_{21}(k,z)$ during cosmic dawn on the sum of neutrino masses $M_\nu$ and the neutrino self-interaction coupling strength $\log_{10}G_{\rm eff}\,[{\rm MeV^{-2}}]$ in the case of strong (left), moderate (center) and mild interaction (right). The analysis has been run on the full set of parameters listed in Tables~\ref{tab:fiducial_cosmo} and \ref{tab:fiducial_astro}, and then marginalizing with uninformative priors over all but the two parameters shown. The 21-cm measurement from HERA can provide stronger constraints over CMB-S4 for the MI$\nu$ model, though our projections for CMB-S4 are conservative. For the mI$\nu$ model, the individual experiments have poor constraining power, but the joint analysis of HERA and CMB-S4 breaks the degeneracy between $\log_{10}G_{\rm eff}$ and $M_\nu$, providing a significant improvement. See also Table~\ref{tab:results}.}
    \label{fig:moneyplot}
\end{figure*}

Our goal in this work is to understand if and how well upcoming 21-cm and CMB data can probe models of self-interacting massive neutrinos.
The main results of our Fisher analyses are summarized in Table~\ref{tab:results}: we provide forecasts for the $2\sigma$ uncertainty on the self-interaction coupling strength $\log_{10}(G_{\rm eff}/{\rm MeV}^{-2})$ for the three self-interacting neutrino models of interest, using the popII astrophysical model in Table~\ref{tab:fiducial_astro} as our baseline for the cosmic dawn analysis.
Figure~\ref{fig:moneyplot} shows the forecasts for $M_\nu$ and $\log_{10} G_{\rm eff}$, marginalizing over all other cosmological and astrophysical parameters, for CMB and 21-cm cosmic dawn.

The full triangle plot for mI$\nu$ with all cosmological and astrophysical parameters for the analyses involving 21-cm cosmic dawn is shown in Fig.~\ref{fig:all_ellipses} in Appendix~\ref{sec:contours}.
The full triangle plots for mI$\nu$ with all cosmological parameters for the Lunar D and LRA1 analyses are shown in Fig.~\ref{fig:all_ellipses_DA} in Appendix~\ref{sec:contours}.
All triangle plots in Appendix~\ref{sec:contours} also include the joint analysis with CMB-S4.

\vspace*{-.2cm}
\subsubsection{CMB-S4 forecasts}
\vspace*{-.1cm}

We have validated our pipeline by running the Fisher matrix analysis for the $\Lambda$CDM+$N_{\rm eff}+M_\nu$ cosmology, with no self-interacting neutrinos. Our results on the neutrino parameters are broadly in agreement with forecasts in the CMB-S4 Science Book~\cite{Abazajian:2019eic}. 

Regarding the constraints on $\log_{10}G_{\rm eff}$ for self-interacting neutrinos, Ref.~\cite{Das:2023npl} performed a similar forcast analysis. However, that work considered three massless self-interacting neutrinos with a fiducial of $\log_{10}G_{\rm eff}= -1.92$, matching the best fit obtained in Ref.~\cite{Das:2020xke} using {\it Planck} 2018 TT+TE+EE+lowE data. Compared to our work here, Ref.~\cite{Das:2020xke} adopts a slightly different definition of $\log_{10}G_{\rm eff}$, which is in line with Ref.~\cite{Cyr-Racine:2013jua}. The noise treatment is also slightly different from ours. Their forecast uncertainty is $2\sigma_{\log_{10}G_{\rm eff}} \simeq 10\%$, close to our value in Table~\ref{tab:fiducial_cosmo}. We have verified that assuming $M_\nu =0$ and changing the $\log_{10}G_{\rm eff}$ fiducial value and noise accordingly, the difference between our result and theirs reduces. 

Reference~\cite{Das:2020xke} also found an upper limit on a moderate interaction at $\log_{10}G_{\rm eff}= -4.05$, with $2\sigma_{\log_{10}G_{\rm eff}}= 50\%$ from \textit{Planck}+ACT data~\cite{ACT:2020gnv}. This result is consistent with ours for MI$\nu$ from CMB-S4 alone.
For the smaller interaction of the mI$\nu$ model, our result indicates that the bound from CMB-S4 would be very weak.  

\vspace*{-.3cm}
\subsubsection{HERA forecasts}\label{sec:HERA_forecasts}
\vspace*{-.1cm}

HERA provides better sensitivity over CMB-S4 to $\log_{10}G_{\rm eff}$ for the MI$\nu$ model.
The constraining power arises from the indirect sensitivity to the matter power spectrum at small scales: the enhancement bump increases the abundance of the small- and intermediate-mass halos that host the galaxies that dominate the evolution of the 21-cm signal during cosmic dawn.
This result opens the possibility of exploring a new region of the parameter space, which so far has been inaccessible using CMB primary anisotropies and galaxy surveys.

Although HERA also outperforms CMB-S4 in constraining $\log_{10}G_{\rm eff}$ for the mI$\nu$ model, the right panel of Fig.~\ref{fig:moneyplot} shows that HERA is limited by the anti-correlation with $M_\nu$. {This strong degeneracy arises from the very similar shapes of the halo mass function predicted by $M_\nu$ and $G_{\rm eff}$ in the halo mass range $M_h \in 10^{10}-10^{11}, M_\odot$, which is particularly relevant for the 21-cm signal (as can be seen in Fig. 5):}
increasing $\log_{10}G_{\rm eff}$ makes it easier to form halos in the relevant $M_h$ range, which speeds up the evolution of $T_{21}(z)$ and leads to a smaller $\Delta_{21}^2(k,z)$ for $k\sim[0.1,0.5]\,{\rm Mpc^{-1}}$, where HERA is sensitive.\footnote{It may seem counterintuitive that a larger HMF leads to a smaller $\Delta_{21}^2(k,z)$. Recall that the amplitude of the 21-cm signal depends not just on the abundance of the astrophysical sources, but also on the interplay of their radiation fields, the way they heat and ionize the gas, etc. In particular, having more galaxies at $z=13$ implies producing more X-rays, so $T_{\rm gas}$ and $T_s$ get closer to the CMB temperature, thereby decreasing $T_{21}$.} The same effect would still occur if $M_\nu$ were smaller than our fiducial choice, thus explaining the anti-correlation. In other words, lighter neutrinos would compensate for the effect of mI$\nu$ on $\Delta^2_{21}(k,z)$. {However, the shape of the degeneracy changes with different fiducial values, as illustrated in the other panels of the figure. In fact, different values of $G_{\rm eff}$ lead to different shapes of the matter power spectrum on the scales relevant to the evolution of the 21-cm signal. 
}

A very interesting possibility to further improve the constraining power is to combine 21-cm and CMB datasets.
Table~\ref{tab:results} highlights the great potential that joint CMB and 21-cm cosmic dawn analyses have for self-interacting massive neutrinos.
Under all three self-interaction models, the joint analysis reduces the uncertainty on $\log_{10}G_{\rm eff}$ to $\sim 10\%$.
As shown in Fig.~\ref{fig:moneyplot}, the combination of CMB and 21-cm is crucial for breaking the degeneracy between neutrino parameters, as well as other degeneracies that may exist within the parameter set, as highlighted in Fig.~\ref{fig:all_ellipses}.
Incorporating 21-cm data from the cosmic dawn helps mitigate the degeneracy between $M_\nu$, $\tau$, and $A_s$; in this context, we have verified that our analysis returns a comparable result with the analysis in Ref.~\cite{Shmueli:2023box}. 
There is also a slight improvement of the constraining power on the astrophysical parameters.

Thus far, the discussion of our results has focused on our baseline astrophysical scenario, where only popII stars are included. We now consider what happens when we include popIII stars in the analysis. {Including popIII stars in the analysis generally produces more conservative results, since they increase the number of parameters and the uncertainties in the astrophysical model. Moreover, they anticipate the onset of cosmic dawn and reionization, thus shifting the 21-cm signal to higher $z$ and decreasing the overall amplitude of its power spectrum in the HERA observational band, see Fig.~\ref{fig:app_astro}. However, the existence of popIII stars may increase the relative contribution of small-scale structures to the evolution of the 21-cm signal, hence improving our capability to probe them. In fact,}
having been formed in mini-halos, popIII stars indirectly probe the small-scale regime, $k\gtrsim 30\,{\rm Mpc^{-1}}$ (compare with Fig.~\ref{fig:pm_to_hmf}). In the mI$\nu$ and MI$\nu$ models, these scales are boosted; hence, the number of mini-halos is strongly enhanced, anticipating the onset of cosmic dawn and increasing the amplitude of the 21-cm power spectrum at high $z$. Meanwhile, the suppression of small scales in the SI$\nu$ model weakens the amplitude of the 21-cm power spectrum; this suppression, in addition to the larger number of parameters needed to analyze the popII+popIII scenario, worsens the constraining power of $\Delta_{21}^2(k,z)$.

\begin{figure*}[t!]
    \centering
\includegraphics[width=.75\columnwidth]{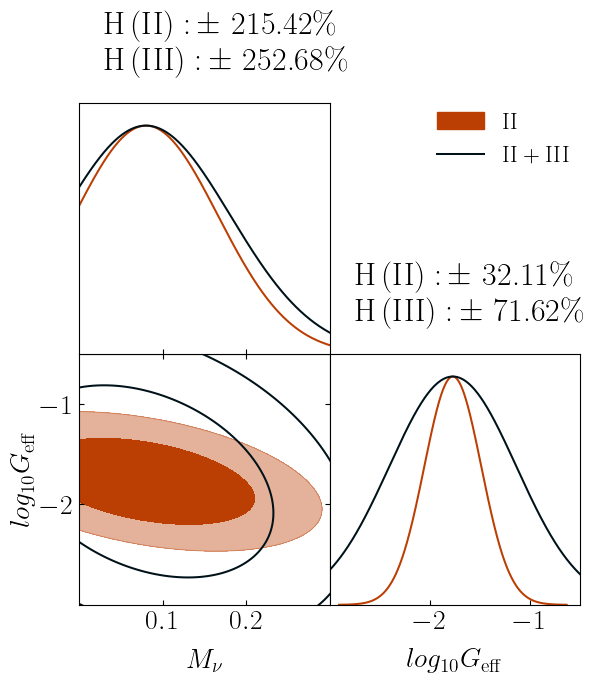}
\includegraphics[width=.75\columnwidth]{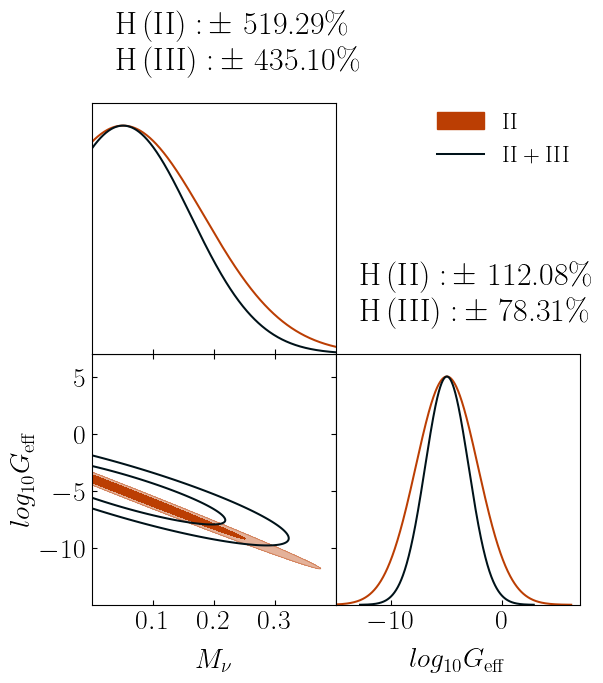}
\caption{Forecasts for the constraining power of HERA for the SI$\nu$ (left) and mI$\nu$ (right) models, using the only-popII (red) and popII+popIII (black) scenarios. The numbers above each panel indicate $2\sigma$ constraints.}
    \label{fig:pop3_ellipses}
\end{figure*}

To help visualize this point, we compare the only-popII and popII+popIII scenarios in Fig.~\ref{fig:pop3_ellipses}, where the HERA results for only-popII are the same from Fig.~\ref{fig:moneyplot}.
The enhanced small-scale power in the mI$\nu$ model helps to relax the degeneracy between $M_\nu$ and $\log_{10}G_{\rm eff}$, improving the constraining power on the latter; the $2\sigma_{\log_{10}G_{\rm eff}}$ constraints from Table~\ref{tab:results}, in fact, shift from $112\%$ to $78\%$. In contrast, the constraints on the SI$\nu$ model are weaker when popIII stars are included in the analysis, with $2\sigma_{\log_{10}G_{\rm eff}}$ shifting from $32\%$ to $72\%$. {Further improvements may come from external constraints on the popIII star formation rate, which would help adding priors to the astrophysical parameters. Despite not yet achieved, a direct observation of popIII stars is in the reach of current and upcoming high redshift experiments, e.g.,~JWST, for which candidate popIII galaxies have been already observed and used for a first calibration of the luminosity function~\cite{Fujimoto:2025kbv}. The synergy between these surveys and 21-cm studies will surely foster our knowledge on these elusive sources.}

Finally, we note that variations in the fiducial astrophysical model may vary the evolution of the 21-cm signal (see, e.g.,~Refs.~\cite{Park:2018ljd,Mesinger:2016ddl,Cohen:2016jbh}). This may affect our final results, since the 21-cm power spectrum would shift in redshift, with cosmic dawn being either anticipated or delayed, hence changing the sensitivity of HERA to its amplitude on the different scales. However, since none of these parameters can introduce scale-dependent effects, we expect the constraining power on $G_{\rm eff}$ to remain in the same ballpark. For example, a smaller value of $f_{*,7}$, the popIII star formation efficiency, would lead to $\sigma_{\log_{10}G_{\rm eff}}$ in between the values we quoted for the two scenarios.

\vspace*{-.2cm}
\subsubsection{Lunar array forecasts}

Lunar detectors can probe neutrino properties more directly: there are no astrophysical sources during the dark ages and thus the 21-cm signal is shaped by only the cosmological parameters. Their great advantage comes from their capability of accessing very small scales, which could be of particular interest for the MI$\nu$ and mI$\nu$ models.
As shown in Fig.~\ref{fig:Pm}, these models feature a boost in $P_m(k,z)$ for $k\in[0.1,10]\,{\rm Mpc}^{-1}$. While these scales are outside the reach of Lunar D, they do fall within some of the redshift bins of LRA1.

\begin{figure*}[t!]
    \centering
\includegraphics[width=.75\columnwidth]{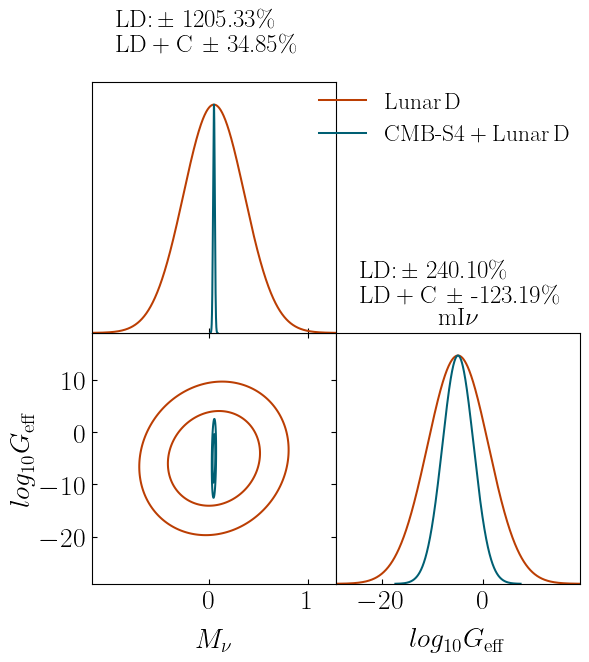}
\includegraphics[width=.75\columnwidth]{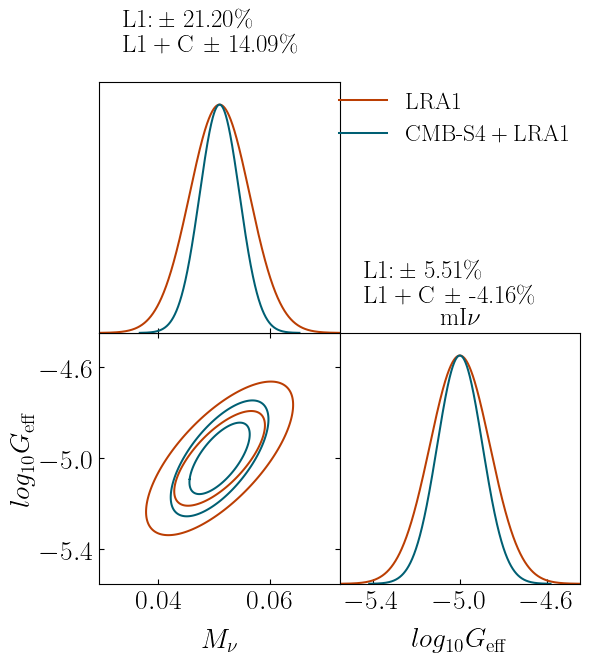}
\vspace*{-.2cm}
\caption{Confidence ellipses on the neutrino parameters for Lunar D (left) and LRA1 (right) for the mI$\nu$ model. We consider the experiments alone (red) or in combination with CMB-S4 (blue). The numbers above each panel indicate $2\sigma$ constraints.}
    \label{fig:nu_ell_DA}
\end{figure*}

We show in Fig.~\ref{fig:nu_ell_DA} the confidence ellipses for the mI$\nu$ model on the neutrino parameters $M_\nu$ and $\log_{10}G_{\rm eff}$; the two panels correspond  to the Lunar D and LRA1 experimental configurations, respectively. These plots, together with the 
results in Table~\ref{tab:results}, show that LRA1 could reach a strong detection of neutrino self-interactions, even for quite small values of the coupling strength. The ability of LRA1 to reach very small scales is the key to constraining not only $\log_{10}G_{\rm eff}$, but all the other cosmological parameters as well; this is evident by looking at the confidence ellipses on the full set of parameters in Fig.~\ref{fig:all_ellipses_DA}. 

Comparing the performances of the two experiments in Figs.~\ref{fig:nu_ell_DA} and~\ref{fig:all_ellipses_DA}, 
in fact, we notice that the confidence ellipses in the LRA1 case show fewer degeneracies among parameters (e.g.,~between $h$ and $\Omega_c$ or between $M_\nu$ and $A_s$), which leads to better constraints on the full parameter set. For the same reason, combining LRA1 with CMB-S4 does not improve the constraining power on $\log_{10}G_{\rm eff}$ by much. Using CMB-S4 is instead helpful in breaking the internal degeneracies in the Lunar D case, improving its capability of constraining the MI$\nu$ scenario.

{In our dark ages results so far, we evolved the matter density field in \texttt{21cmFirstCLASS} linearly at $z > 35$ to reduce computational cost. However, this approximation neglects the mode coupling induced by the baryon–dark matter relative velocity, $v_{\rm cb}$, which modulates small-scale power on large scales. As demonstrated in Ref.~\cite{Ali-Haimoud:2013hpa}, this non-linear evolution leads to a small enhancement of the 21-cm power spectrum on scales $k \in [0.005\,{\rm Mpc}^{-1}\, 1\,{\rm Mpc}^{-1}]$, caused by the coupling of power from smaller scales, $k \gtrsim 30\,{\rm Mpc}^{-1}$.

\medskip
According to Ref.~\cite{Ali-Haimoud:2013hpa}, in $\Lambda$CDM this effect becomes non-negligible (contributing up to $\sim$10\% compared to the purely linear case) only at large scales ($k \lesssim 0.01\,{\rm Mpc}^{-1}$) and low redshifts ($z \sim 30$), which are only marginally within the $k$ and $z$ ranges probed by the interferometers considered in this work. 
In the I$\nu$ models analyzed, the relevance of this effect may be slightly modified. As shown in Fig.~\ref{fig:Pm}, both SI$\nu$ and MI$\nu$ suppress power on scales $k \gtrsim 30\,{\rm Mpc}^{-1}$, further diminishing the impact of $v_{\rm cb}$. In contrast, the mI$\nu$ model shows enhanced small-scale power, which, once processed through the mode coupling introduced by $v_{\rm cb}$, could lead to a modest increase in the signal observed by Lunar D. This may slightly boost the SNR and help reduce the strong degeneracies discussed earlier. Nevertheless, we do not expect this to alter our conclusions, especially when the 21-cm power spectrum is combined with CMB data.
}


\vspace*{-.4cm}
\section{Final remarks}\label{sec:discussion}
\vspace*{-.1cm}

Alongside traditional methods of studying cosmology, such as CMB observations and galaxy surveys, the exploration of the 21-cm signal from cosmic dawn and the dark ages is expected to become increasingly important in the coming years. This observable will provide a powerful tool, offering insights into cosmological models beyond $\Lambda$CDM and extensions to the Standard Model of particle physics.

In this paper, we investigate self-interacting massive neutrinos.
We consider the scale-dependent changes in the matter power spectrum that self-interacting neutrinos introduce and how those changes propagate to the 21-cm power spectrum.
Scales that enter the horizon before neutrino self-decoupling are damped, while modes entering the horizon around the time of self-decoupling are strongly boosted. The scales on which the boost occurs depend on the coupling strength of the self-interaction, $G_{\rm eff}$.

We study three fiducial models: massive neutrinos with a strong (SI$\nu$), moderate (MI$\nu$), and mild (mI$\nu$) self-interaction.
We explore two astrophysical scenarios for the 21-cm signal from cosmic dawn: one in which star formation is driven solely by population II stars within atomic cooling galaxies, and another that also includes the effect from population III stars within molecular cooling galaxies, associated with high-redshift mini-halos.
Finally, we perform Fisher forecasts to analyze various experimental setups involving separate and joint analyses for CMB and 21-cm experiments.
We consider CMB-S4, HERA for the 21-cm cosmic dawn signal, and two Lunar configurations for the 21-cm dark ages signal.

We find that the 21-cm signal during the cosmic dawn provides an indirect
way to investigate physics that impacts small-scale structure, due to the interplay between the halo mass function (whose shape depends on the matter power spectrum) and the star formation rate density. Galaxies that mainly drive cosmic dawn evolution are hosted by halos in the $10^9-10^{11}\,M_\odot$ mass range, whose abundance is determined by $P_m(k,z)$ on $k\sim 1-100\,{\rm Mpc}^{-1}$. This range of scales, that are just beyond the reach of current CMB observations, is precisely where neutrino self-interactions could play a role.

A key finding in our work is that joint CMB and 21-cm analyses can provide a substantial improvement in constraining power over an individual analysis of either alone.
The CMB and 21-cm power spectra access different cosmic scales, so together they are a powerful probe of correlated physics that span a wide range in $k$.
Their combination can break internal degeneracies among parameters, allowing one probe to tightly constrain neutrino properties more directly, while the other probe helps by constraining other parameters in the joint analysis interchangeably. For example, as Refs.~\cite{Liu:2015txa,Shmueli:2023box} have shown in the context of $\Lambda$CDM, the 21-cm signal can be used to mitigate the $M_\nu-\tau-A_s$ degeneracy in CMB studies; in our case, it also leads to improving the constraints on the self-interaction.

During the dark ages, the 21-cm power spectrum carries pristine information about cosmology. Even if challenging, we find that observing the 21-cm signal in this epoch can provide very strong constraints on the cosmological parameters that affect the shape of the matter power spectrum.

To conclude, our results highlight the unique potential of the 21-cm power spectrum in probing new physics beyond $\Lambda$CDM and beyond the SM. By complementing traditional observables such as the CMB, 21-cm measurements can play a key role in constraining self-interacting massive neutrinos, as well as other beyond-SM scenarios, shedding light on fundamental physics on previously inaccessible scales.

\vspace*{-.5cm}
\begin{acknowledgments}
The authors thank Jordan Flitter and Gali Shmueli for useful and enlightening discussions. We acknowledge the efforts of the  {\tt 21cmFAST}, {\tt 21cmFirstCLASS}, \texttt{21cmSense}, and {\tt CLASS} authors to produce state-of-the-art public 21-cm and CMB codes. SL thanks Jose Louis Bernal and the Instituto de Física de Cantabria for hospitality during the last stages of this work and for useful discussions.

SL is supported by an Azrieli International Postdoctoral Fellowship. EDK acknowledges joint support from the U.S.-Israel Bi-national Science Foundation (BSF, grant No.\,2022743) and the U.S. National Science Foundation (NSF, grant No.\,2307354), and support from the ISF-NSFC joint research program (grant No.\,3156/23).
SG and KB acknowledge support from the NSF under Grant No.~PHY-2413016.
AR acknowledges funding from the Italian Ministry of University and Research (MUR) through the ``Dipartimenti di eccellenza'' project ``Science of the Universe'', and thanks the University of Texas, Austin, for hospitality in the first stages.

\end{acknowledgments}


\appendix
\section{Modified Boltzmann equations for neutrino self-interaction}
\label{sec:modboltz}
In this section, we describe the modifications of the Boltzmann equations implemented in \texttt{CLASS} which we have made publicly available in the code \texttt{nuCLASS}. We sketch our modifications, which follow Ref.~\cite{Kreisch:2019yzn}. 

The low-energy neutrino interaction Lagrangian in the presence of a scalar mediator $\varphi$ of mass $m_\varphi$ is given by
\begin{equation}
    \label{eq:nulag}
    \mathcal{L} \supset g_{\nu,ij}\varphi\bar{\nu}_i\nu_j\;,
\end{equation}
where $\nu_i$ is the left-handed neutrino spinor of flavor $i$. In this work, we consider flavor-universal interactions and set the coupling
\begin{equation}
    \label{eq:flavorunivdef}
    g_{\nu,ij} = g_\nu\delta_{ij}\;,
\end{equation}
where $\delta_{ij}$ is the Kronecker delta function and $g_\nu$ is the universal coupling. The scattering matrix element for 4-neutrino scattering in the presence of the interaction is
\begin{eqnarray}
    \label{eq:modmsq}
    |\mathcal{M}|^2 = \sum_{\rm spin} |\mathcal{M}|^2_{\nu_i + \nu_j \to \nu_k + \nu_l} \\
    = 2G_{\rm eff}^2(s^2+t^2+u^2)\;,
\end{eqnarray}
where $s,t$ and $u$ are the Mandelstam variables and $G_{\rm eff} = g_\nu^2/m_\varphi^2$ as in Eq.~\eqref{eq:geffdef}. 

The perturbed neutrino distribution function is
\begin{equation}
    \label{eq:Psi_def}
    f_\nu(\mathbf{x},\mathbf{p},\tau) = f^{(0)}_\nu(p,\tau)[1+\Psi(\mathbf{x}, \mathbf{p},\tau)]\;,
\end{equation}
where $\mathbf{x}$ is a spatial coordinate, $\mathbf{p}$ is the proper momentum with magnitude $p = |\mathbf{p}|$, and $\tau$ is conformal time. The background neutrino distribution function is taken to be Fermi-Dirac:
\begin{equation}
    \label{eq:FDnu}
    f^{(0)}_\nu(p,\tau) = \dfrac{1}{e^{p / T_\nu} + 1}\;,
\end{equation}
where $T_\nu$ is the neutrino temperature. The perturbation $\Psi(\mathbf{k}, \mathbf{p},\tau)$ encodes the inhomogeneities in the neutrino distribution function in the conjugate-momentum space, and it is expanded into multipole moments in the following manner~\cite{Ma:1994dv}:
\begin{equation}
    \label{eq:Psi_l}
    \Psi(\mathbf{k}, \mathbf{p},\tau) = \sum_{\ell = 0}^\infty (-i)^\ell (2\ell +1) \Psi_\ell(k,p,\tau) P_\ell(\mu)\;,
\end{equation}
where $P_\ell(\mu)$ is the Legendre polynomial with $\mu = \hat{k}\cdot\hat{p}$.
The momentum-dependent Boltzmann equation in the presence of the interaction is
\begin{multline}
    \dot{\Psi}_\ell
    + k\dfrac{q}{\epsilon}\left(\dfrac{\ell +1}{2\ell +1}\Psi_{\ell+1} - \dfrac{\ell}{2\ell +1}\Psi_{\ell -1}\right) \\
    +\left(\dot{\phi}\delta_{\ell 0} + \dfrac{k\epsilon}{3 q}\psi\delta_{\ell1}
    \right)\dfrac{d\ln f_\nu^{(0)}}{d\ln q} \\ = -a G_{\rm eff}^2 T_\nu^5\left(\dfrac{1}{f_\nu^{(0)}}\dfrac{T_{\nu,0} }{ q}\right) \times \\
    \left[A\left(
    q \over T_{\nu,0}\right) + B_\ell\left(
    q \over T_{\nu,0}\right) - 2D_\ell\left(
    q \over T_{\nu,0}\right)\right]\Psi_\ell\;,
\end{multline}
where the overdot $(\dot{~})$ denotes a derivative with respect to $\tau$, $q \equiv ap$ is the comoving momentum, $\epsilon = \sqrt{q^2 + a^2m_\nu^2}$, $m_\nu$ is the neutrino mass, $\phi$ and $\psi$ are Newtonian potentials, $a$ is the scale factor, and $T_{\nu,0}$ is the temperature of neutrinos today ($a=1$). Due to the conservation of number density (i.e., detailed balance) for elastic scattering, we have
\begin{equation}
    A(x) + B_0(x) - 2D_0(x) = 0\;,
\end{equation}
where $x \equiv q/T_{\nu,0}$. Since neutrinos self-scatter, there is a conservation of total momentum in the neutrino sector, which results in 
\begin{equation}
   A(x) + B_1(x) - 2D_1(x) = 0\;.
\end{equation}
We precompute the combination 
\begin{equation}
    S_l(x) = \dfrac{1}{x f_\nu^{(0)}}
    \left[A(x) + B_\ell(x) - 2D_\ell(x)\right]\;,
\end{equation}
for different $x$ and $\ell$ values and store the results as an interpolation file to be used for \texttt{nuCLASS} computations.

The functions $A$, $B_\ell$, and $C_\ell$ associated with the collision term are as follows:
\begin{widetext}
\begin{align}
    A(x_1) = \frac{1}{8\pi^3}\int_0^\infty \frac{e^{x_2}dx_2}{e^{x_2}+1}&\int_0^{x_1+x_2}\frac{x_3dx_3}{(e^{x_3}+1)(e^{x_1 +x_2 -x_3}+1)}\int_{{\rm Max}[\eta_-,1]}^1\frac{|\bar{\mathcal{M}}_\eta(x_1,x_2,x_3,\eta)|^2}{\sqrt{x_1^2 + x_3^2 - 2x_1x_3\eta}}d\eta\;,\\
    B_\ell(x_1) = \frac{1}{8\pi^3 x_1}\int_0^\infty \frac{x_2^2 e^{x_2}dx_2}{e^{x_2}+1}&\int_0^{x_1+x_2}\frac{dx_3}{(e^{x_3}+1)(e^{x_1 +x_2 -x_3}+1)}\int_{{\rm Max}[\rho_-,1]}^1\frac{|\bar{\mathcal{M}}_\rho(x_1,x_2,x_3,\rho)|^2 P_\ell(\rho)}{\sqrt{x_1^2 + x_2^2 + 2x_1x_2\rho}}d\rho\;,\\
    D_\ell(x_1) = \frac{e^{-x_1}}{8\pi^3 x_1}\int_0^\infty \frac{dx_2}{e^{x_2}+1}&\int_0^{x_1+x_2}\frac{e^{x_3}x_3^2dx_3}{(e^{x_3}+1)(e^{-(x_1 +x_2 -x_3)}+1)}\int_{{\rm Max}[\eta_-,1]}^1\frac{|\bar{\mathcal{M}}_\eta(x_1,x_2,x_3,\eta)|^2 P_\ell(\eta)}{\sqrt{x_1^2 + x_3^2 - 2x_1x_3\eta}}d\eta\;,
\end{align}
where
\begin{equation}
|\bar{\mathcal{M}}_\eta(x_1,x_2,x_3,\eta)|^2 = 
    \Delta_{2,\eta}\frac{3b_\eta^2-4a_\eta c_\eta}{8a_\eta^2} - \Delta_{1,\eta}\frac{b_\eta}{2a_\eta} + \Delta_{0,\eta}
\end{equation}
\begin{align}
    &a_\eta = -x_2^2(x_1^2 + x_3^2 - 2x_1x_3\eta)\\
    &b_\eta = 2x_2(x_1 -x_2\eta)[x_1 x_2 + x_3(x_1+x_2)+x_1x_3\eta]\\ 
    &c_\eta = - [x_1x_2 -x_3(x_1 + x_2) + x_1x_3\eta]^2+x_2^2x_3^2(1-\eta^2)\\
    &\Delta_{2,\eta} = x_1^2x_2^2,\\
    &\Delta_{1,\eta} = x_1^2x_2(x_3 - 2x_2 - x_3\eta),\\
    &\Delta_{0,\eta} = x_1^2(x_2^2 - x_2x_2 + x_3^2) + x_1x_3 \eta (x_1x_2 - 2x_1x_3 + x_1x_3\eta)\\
    &\eta_- = \frac{(x_1+2x_2)x_3 - 2x_2(x_1+x_2)}{x_1x_3}
\end{align}
\begin{equation}
|\bar{\mathcal{M}}_\rho(x_1,x_2,x_3,\rho)|^2 = 
    \Delta_{2,\rho}\frac{3b_\rho^2-4a_\rho c_\rho}{8a_\rho^2} - \Delta_{1,\rho}\frac{b_\rho}{2a_\rho} + \Delta_{0,\rho}
\end{equation}
\begin{align}
    &a_\rho = -x_3^2(x_1^2 + x_2^2 + 2x_1x_2\rho)\\
    &b_\rho = 2x_3(x_1+x_2\rho)
    [x_2 x_3 + x_1(x_3-x_2)+x_1x_2\rho]\\ 
    &c_\rho = - [x_2x_3 +x_1(x_3 - x_2) + x_1x_2\rho]^2+x_1^2x_3^2(1-\eta^2)\\
    &\Delta_{2,\rho} = x_1^2x_3^2,\\
    &\Delta_{1,\rho} = x_1^2x_3(x_2 - 2x_3 - x_2\rho),\\
    &\Delta_{0,\rho} = x_1^2(x_2^2 - x_2x_2 + x_3^2) + x_1x_2 \rho (x_1x_3 - 2x_1x_2 + x_1x_2\rho)\\
    &\rho_- = \frac{x_1x_2 -2(x_1+x_2)x_3 + 2x_3^2 }{x_1x_2} \; .
\end{align}
\end{widetext}

\section{21-cm signal modeling}\label{sec:app_21cm}

The main observables for studying the 21-cm signal at high $z$ are the brightness temperature $T_{21}(z)$ and its spatial fluctuations, used to estimate the 21-cm power spectrum in Eq.~\eqref{eq:Delta_21}. The former is defined in Eq.~\eqref{eq:T21}, where $T_{\rm CMB}(z)$ is the CMB temperature and $T_s(z)$ the spin temperature.
While the CMB temperature decreases with redshift due to the expansion of the Universe, $T_{\rm CMB}(z)\propto (1+z)$, the evolution of the spin temperature is driven by the coexistence of different processes that excite H in the intergalactic medium (IGM). In thermal equilibrium, we can write
\begin{equation}\label{eq:Ts}
    T_s^{-1}(z)=\frac{x_{\rm CMB}T_\gamma^{-1}(z)+x_{\rm coll}T_k^{-1}(z)+x_{\alpha}T_\alpha^{-1}(z)}{x_{\rm CMB}(z)+x_{\rm coll}(z)+x_{\alpha}(z)},
\end{equation}
where $T_k(z)$ is the kinetic temperature of the H gas in the IGM, and $T_\alpha(z)\simeq T_k(z)$~\cite{Field:1959ApJ...129..536F} is the color temperature of Lyman-$\alpha$ photons (Ly$\alpha$). The evolution of $T_k(z)$ over cosmic time is found by solving the following differential equation~\cite{Pritchard:2006sq}:
\begin{equation}\label{eq:Tk}
\begin{aligned}
\frac{dT_k}{dz} &= {\frac{dt}{dz}\biggl[\frac{2T_k}{3n_b}\frac{dn_b}{dt}+\frac{2}{3k_B(1+x_e)}\sum_j{\epsilon_j}
-\frac{T_k}{1+x_e}\frac{dx_e}{dt}\biggr]}\\
&{=\frac{dt}{dz}\biggl[\Gamma_C(T_\gamma-T_k)+\frac{dT_k}{dz}\biggl|_{\rm CMB}+\frac{dT_k}{dz}\biggl|_{\rm X-ray}}\\
&\quad{+\frac{2T_k}{1+z}+\frac{2T_k}{3}\frac{D(z)}{\delta_b(z)+1}\frac{dD(z)}{dz}-\frac{T_k}{1+x_e}\frac{dx_e}{dz}\biggr],}
\end{aligned}
\end{equation}
where $dt/dz=-[H(z)(1+z)^{-1}]$, {$n_b$ is the total number density, and $x_e(z)=n_e/(n_{\rm H}+n_{\rm He})$ the free-electron fraction, namely the ratio between the electron number density and the total number density of H and helium (He) nuclei. In the first row, the first term in parenthesis accounts for adiabatic cooling, while the second for the $j$- external sources that contribute with heating (or cooling) rate per baryon $\epsilon_j$. Finally, the last term represents the change in the total number of free-electrons due to ionization. Following Ref.~\cite{Mesinger:2010ne},
in the second and third rows of the equation we explicitly express the evolution of $T_k$ in terms of the adiabatic cooling due to the Universe expansion, the adiabatic cooling due to the Compton heating rate
$\Gamma_C\propto T_\gamma^4x_e/(1+x_e)$, and the external sources of heating (CMB and X-ray).} 

\noindent The evolution of $x_e(z)$ in turn can be found by solving the coupled differential equation~\cite{Peebles:1968ja}
\begin{equation}\label{eq:xe}
    \frac{dx_e}{dz} =\frac{dt}{dz}\left[\frac{dx_e}{dt}\biggl|_{\rm reio}+\mathcal{C}(\beta_{\rm ion}(1-x_e)-\alpha_{\rm rec}n_{\rm H}x_e^2)\right],
\end{equation}
where ${dx_e}/{dt}|_{\rm reio}$ is the reionization rate at late times, $\mathcal{C}$ represents the Peebles coefficient,\footnote{The interested reader can read e.g.,~App.~B in Ref.~\cite{Flitter:2023rzv} for discussion on the Peebles coefficients.} while $\alpha_{\rm rec}$ is the recombination rate and $\beta_{\rm ion}$ the early photoionization rate.

In Eq.~\eqref{eq:Tk}, changes in $T_k$ can be sourced by either heating due to external sources (second line), or by adiabatic cooling/heating (first line), due to the expansion of the Universe (first term), the clustering of matter (second term) and variations in the total baryon number density sourced by reionization/recombination (third term). 

Among the external sources, we consider the heating due to absorptions of CMB photons~\cite{Ma:1995ey,Seager:1999bc,Seager:1999km,Barkana:2000fd,Naoz:2005pd} and X-rays from astrophysical sources during cosmic dawn. The contribution of the X-ray heating can be estimated based on the X-ray flux~\cite{Pritchard:2006sq}
\begin{equation}
    J_X(\boldsymbol{x},\nu) = \int dR \dot{\rho}_*(\boldsymbol{x},R)\epsilon_X(\nu')e^{-\tau_X}(\nu',R),
\end{equation}
where $\boldsymbol{x}$ is the point (at redshift $z$) at which the flux is computed and $\nu$ the rest-frame frequency of the X-ray radiation. The integral over shells of comoving radius $R$ accounts for radiation sourced from different redshifts $z'$, so that the frequency at $\boldsymbol{x}$ is $\nu'=\nu(1+z'(R))/R$. The emissivity $\epsilon_X(\nu')$ accounts for the spectral energy distribution in the X-ray band; its value can be related to the value of the $L_{X,10}/{\rm SFR}$ parameter introduced in Sect.~\ref{sec:astro}. Finally, $\dot{\rho}_*$ is the star formation rate density discussed in Sect.~\ref{sec:21cm_CD}, while $\tau_X(\nu',R)$ the opacity due to the IGM. 

Finally, in Eq.~\eqref{eq:Ts}, the coefficients $x_{\rm CMB}(z)$, $x_{\rm coll}(z)$, $x_\alpha(z)$ represent, respectively, the coupling with CMB (due to the absorption of CMB photons), the collisional coupling (due to collisions between H atoms) and the Ly$\alpha$ coupling (due to the absorption of Ly$\alpha$ photons under the Wouthuysen-Field effect~\cite{Wouthuysen:1952AJ.....57R..31W,Field:1958PIRE...46..240F,Hirata:2005mz}). On one side, $x_{\rm CMB}(z)=(1-e^{\tau_{21}(z)})/\tau_{21}(z)$, which is $\simeq 1$ until reionization begins; on the other, $x_{\rm coll}(z)$ and $x_{\alpha}(z)$ are described by more complicated expressions, which require knowledge of the particle number density (for $x_{\rm coll}$) or of the Ly$\alpha$ flux (for $x_{\alpha}$). The latter, in particular, can be estimated as a function of the star formation rate density~\cite{Barkana:2004vb}, as
\begin{equation}
J_\alpha(\boldsymbol{x},z)=\frac{(1+z)^2}{4\pi}\int dR\dot{\rho}_*\epsilon_\alpha(\nu'),
\end{equation}
where $\epsilon_\alpha(\nu')$ indicates the spectral energy distribution in the Ly$\alpha$ to continuum regime. 

Extensive review on the 21-cm signal can be found in Refs.~\cite{Barkana:2000fd,Furlanetto:2006jb,Pritchard:2011xb}.

\section{Contour plots}
\label{sec:contours}

\begin{figure*}[ht!]
    \centering
    \includegraphics[width=2\columnwidth]{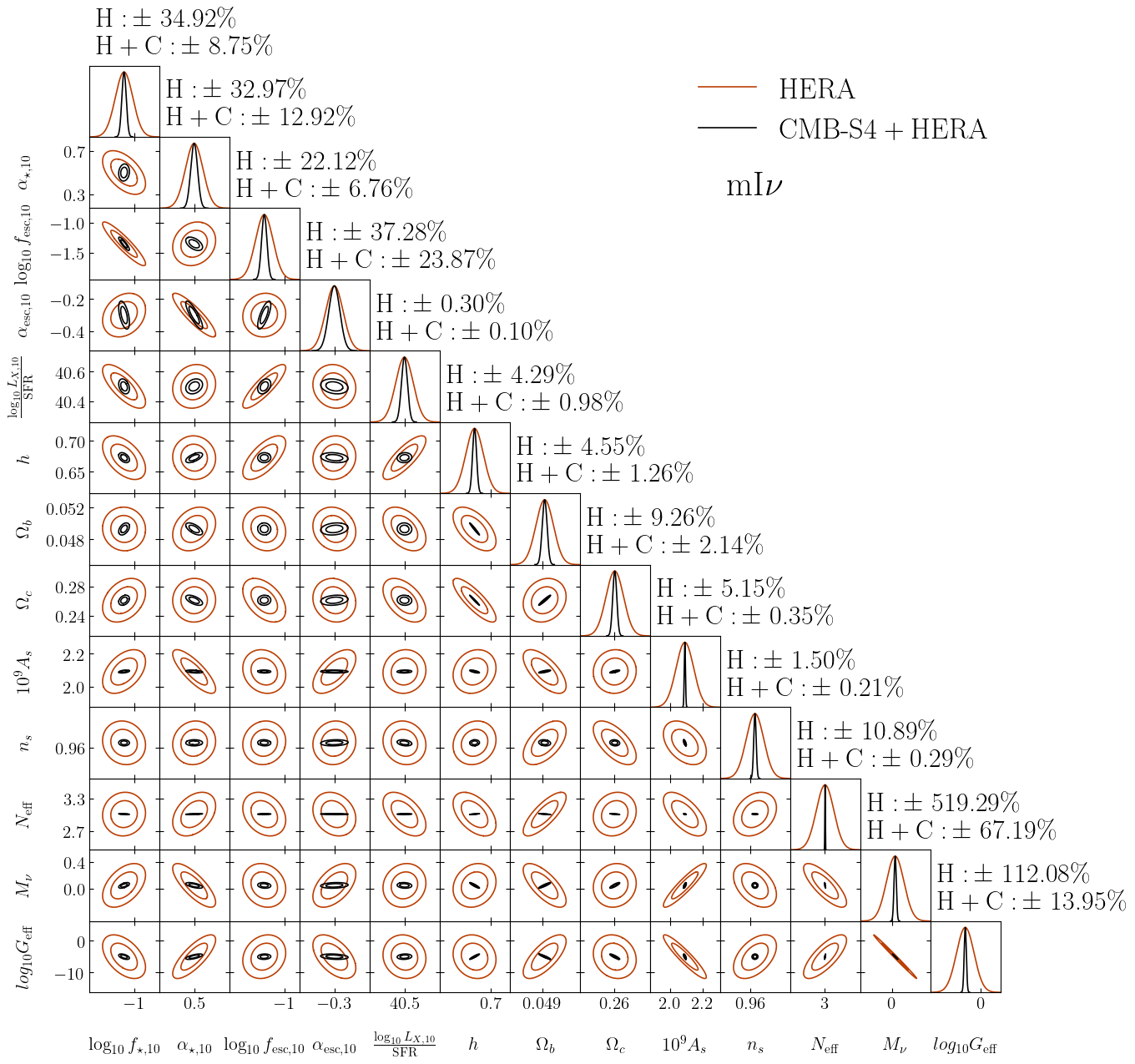}
    \caption{Confidence ellipses and $2\sigma$ forecast uncertainties for all cosmological and astrophysical parameters in the HERA (H, red) and HERA+CMB-S4 (H+C, black) analysis, for the mI$\nu$ model. }
    \label{fig:all_ellipses}
\end{figure*}

\begin{figure*}[h!]
    \centering
    \includegraphics[width=1.3\columnwidth]{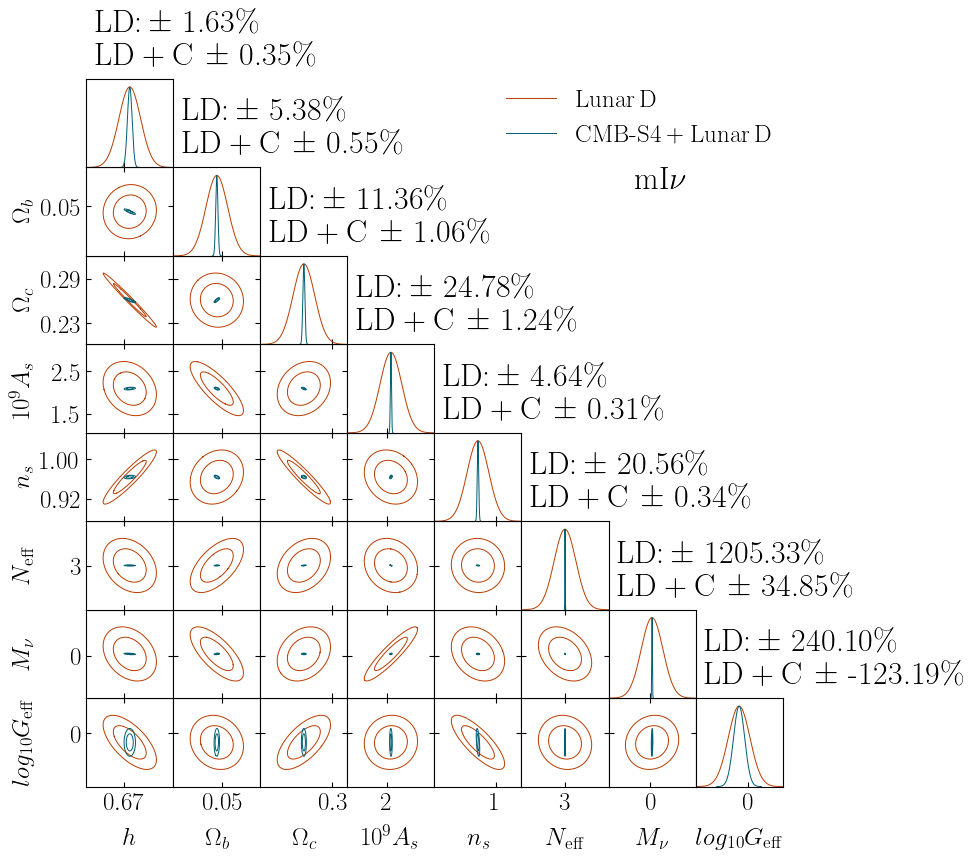}\\
    \includegraphics[width=1.3\columnwidth]{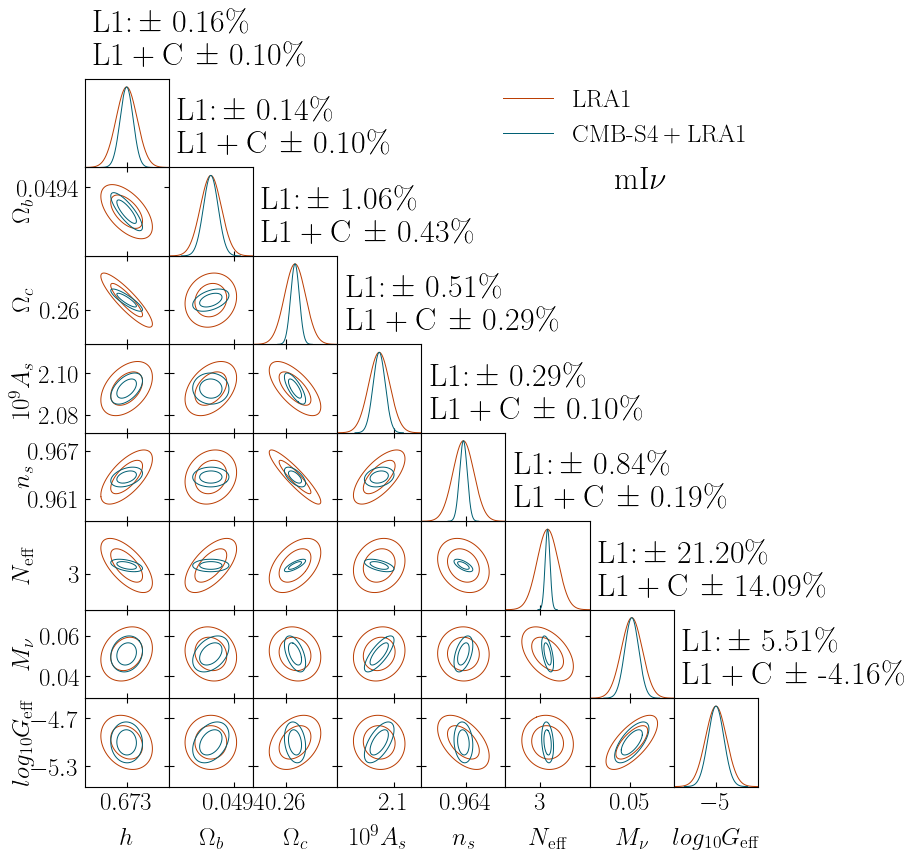}
    \caption{Confidence ellipses and $2\sigma$ forecast uncertainties for the cosmological parameters in the mI$\nu$ model, using Lunar D (top) and LRA1 (bottom), alone (red) or in combination with CMB-S4 (black).}
    \label{fig:all_ellipses_DA}
\end{figure*}

We show in Figs.~\ref{fig:all_ellipses} and~\ref{fig:all_ellipses_DA} the 1$\sigma$ and 2$\sigma$ confidence contours obtained from our analysis in Sec.~\ref{sec:forecast}.
We include these triangle plots to highlight any potential degeneracy between cosmological and astrophysical parameters. To be concise, we only show the mI$\nu$ model, since this is the case that requires 21-cm to explore new regions of parameter space beyond the capabilities of CMB experiments.

Note that $\tau$ is not included in the figures.
Its uncertainty cannot be estimated from the 21-cm signal. In the case of cosmic dawn, we derive forecast constraints on $\tau$ as a function of the other parameters; see Eq.~\eqref{eq:fisher_total} and the related discussion.

\onecolumngrid
\twocolumngrid


\clearpage
\bibliography{biblio.bib}

\end{document}